\documentclass[sigconf]{acmart}

\AtBeginDocument{%
  }

\setcopyright{acmlicensed}
\copyrightyear{2025}
\acmYear{2025}
\acmDOI{10.1145/3706598.3713276}

\acmConference[CHI '25]{Proceedings of the CHI Conference on Human Factors in Computing Systems}{April 26 -- May 1, 2025}{Yokohama, Japan}
\acmISBN{978-1-4503-XXXX-X/18/06}

\acmSubmissionID{2578}

\usepackage{hyperref}
\usepackage[capitalise]{cleveref}
\crefname{table}{Tab.}{Tabs.}
\Crefname{table}{Table}{Tables}
\crefname{section}{Sec.}{Secs.}
\Crefname{section}{Section}{Sections}
\Crefname{figure}{Fig.}{Fig.}
\Crefname{figure}{Figure}{Figures}

\usepackage{CJKutf8}

\newcommand{\inlinecode}[1]{\texttt{\detokenize{#1}}}

\usepackage{enumitem}
\setlist{noitemsep,parsep=0pt,partopsep=0pt}

\usepackage{soul}

\newcommand{\remove}[1]{}

\newcommand{\future}[1]{}

\newcommand{\remark}[1]{{}}
\newcommand{\yzc}[1]{}
\newcommand{\xyc}[1]{}
\newcommand{\revised}[1]{{#1}}

\usepackage[group-separator={$,$}]{siunitx}

\usepackage{xspace}

\newcommand{\numVisShort}{71K\xspace}

\newcommand{\numIllusShort}{108K\xspace}

\newcommand{\numClassified}{\num{178659}\xspace}

\newcommand{\numChecked}{\num{27030}\xspace}

\newcommand{\numBook}{\num{12821}\xspace}
\newcommand{\numBookShort}{12K\xspace}

\newcommand{\numBooksLoC}{\num{1979}\xspace}
\newcommand{\numBooksHL}{\num{9335}\xspace}
\newcommand{\numBooksNDL}{\num{1476}\xspace}

\newcommand{\numChaofanYangRelevantImages}{\num{418}\xspace}

\newcommand{\numDunhuangRelevantImages}{\num{12}\xspace}

\newcommand{\numShugeBooks}{\num{31}\xspace}
\newcommand{\numShugeRelevantImages}{\num{5913}\xspace}

\newcommand{\numPreliminaryRelevantImages}{\num{6343}\xspace}

\newcommand{\datasetName}{ZuantuSet\xspace}

\newcommand{\numBookHasGraphic}{\num{3036}\xspace}
\newcommand{\numBookHasVis}{\num{2459}\xspace}

\newcommand{\Fone}{90.1\%\xspace}
\newcommand{\recall}{90.3\%\xspace}
\newcommand{\precision}{90.0\%\xspace}

\newcommand{\itemType}{graphic\xspace}
\newcommand{\itemTypePlural}{graphics\xspace}

\newcommand{\schemaBook}{\inlinecode{book}\xspace}
\newcommand{\schemaImage}{\inlinecode{image}\xspace}
\newcommand{\schemaGraphic}{\inlinecode{graphic}\xspace}

\usepackage{tikz}

\newcommand{\taxon}{\textsl}

\newcommand{\chinese}[1]{\begin{CJK*}{UTF8}{bsmi}{#1}\end{CJK*}}

\newcommand{\ChineseTerm}[3]{\emph{#1} (\chinese{#2}, #3)}

\newcommand{\ChineseTermWithUrl}[4]{\href{#4}{\emph{#1} (\chinese{#2}, #3)}}

\newcommand{\term}[1]{\emph{#1}}

\newcommand{\GuangYuJi}{Guang yu ji\xspace}
\newcommand{\DongTingQinShiZongPu}{Dongting Qin shi zong pu\xspace}
\newcommand{\BenCaoGangMu}{Bencao gangmu\xspace}
\newcommand{\LeiShu}{leishu\xspace}
\newcommand{\SiShuTuKao}{Si shu tu kao\xspace}
\newcommand{\SiShuWuJingDaQuan}{Si shu Wu jing da quan\xspace}
\newcommand{\WangGongZhiTu}{Wang gong zhi tu\xspace}
\newcommand{\JiuBianZongTu}{Jiu bian zong tu\xspace}
\newcommand{\DongXiFenShanTu}{Dong xi fen shan tu\xspace}
\newcommand{\YiYouTaiJiTu}{Yi you Taiji tu\xspace}
\newcommand{\SiShiTu}{Si shi tu\xspace}
\newcommand{\Shiji}{Shiji\xspace}
\newcommand{\SanDaiShiBiao}{San dai shi biao\xspace}
\newcommand{\GaoZuGongChenHouZheNianBiao}{Gao zu gong chen hou zhe nian biao\xspace}
\newcommand{\ZhenJiuDaCheng}{Zhenjiu dacheng\xspace}
\newcommand{\ShiJing}{Shijing\xspace}
\newcommand{\SiShiChuanShouTu}{Si shi chuan shou tu \xspace}

\newcommand{\dbname}{\textit}

\newcommand{\UrlOfDongXiFenShanTu}{https://iiif.lib.harvard.edu/manifests/view/drs:28472486$30i}
\newcommand{\UrlOfGaoZuGongChenHouZheNianBiao}{https://iiif.lib.harvard.edu/manifests/view/drs:19140085$609i}

\begin{document}

\graphicspath{{assets/imgs}}
\title[ZuantuSet: A Collection of Historical Chinese Visualizations and Illustrations]{ZuantuSet: A Collection of\\ Historical Chinese Visualizations and Illustrations}

\author{Xiyao Mei}
\orcid{0009-0002-3127-1519}
\affiliation{%
  \department{National Key Laboratory of General Artificial Intelligence, and School of Intelligence Science and Technology}
  \institution{Peking University}
  \city{Beijing}
  \country{China}
}
\email{xiyao.mei@stu.pku.edu.cn}

\author{Yu Zhang}
\orcid{0000-0002-9035-0463}
\authornotemark[1]
\affiliation{%
  \department{Department of Computer Science}
  \institution{University of Oxford}
  \city{Oxford}
  \country{United Kingdom}
}
\email{yuzhang94@outlook.com}

\author{Chaofan Yang}
\orcid{0000-0002-3627-7357}
\affiliation{%
  \department{National Key Laboratory of General Artificial Intelligence, and School of Intelligence Science and Technology}
  \institution{Peking University}
  \city{Beijing}
  \country{China}}
\email{chaofanyang@pku.edu.cn}

\author{Rui Shi}
\orcid{0000-0003-0304-9080}
\affiliation{%
  \department{Center for Research on Ancient Chinese History, Peking University}
  \institution{Peking University}
  \city{Beijing}
  \country{China}
}
\email{shirui@pku.edu.cn}

\author{Xiaoru Yuan}
\orcid{0000-0002-7233-980X}
\authornote{Corresponding authors}
\affiliation{%
 \department{National Key Laboratory of General Artificial Intelligence, and School of Intelligence Science and Technology}
  \institution{Peking University}
  \city{Beijing}
  \country{China}
}
\affiliation{%
 \department{National Engineering Laboratory for Big Data Analysis and Application}
  \institution{Peking University}
  \city{Beijing}
  \country{China}
}
\affiliation{%
  \institution{PKU-WUHAN Institute for Artificial Intelligence}
  \city{Wuhan}
  \country{China}
  }
\email{xiaoru.yuan@pku.edu.cn}

\renewcommand{\shortauthors}{Mei et al.}
\begin{abstract}
    Historical visualizations are a valuable resource for studying the history of visualization and inspecting the cultural context where they were created.
When investigating historical visualizations, it is essential to consider contributions from different cultural frameworks to gain a comprehensive understanding.
While there is extensive research on historical visualizations within the European cultural framework, this work shifts the focus to ancient China, a cultural context that remains underexplored by visualization researchers. 
To this aim, we propose a semi-automatic pipeline to collect, extract, and label historical Chinese visualizations.
Through the pipeline, we curate \datasetName, a dataset with over \numVisShort visualizations and \numIllusShort illustrations.
We analyze distinctive design patterns of historical Chinese visualizations and their potential causes within the context of Chinese history and culture.
We illustrate potential usage scenarios for this dataset, summarize the unique challenges and solutions associated with collecting historical Chinese visualizations, and outline future research directions.

\end{abstract}




\maketitle

    \section{Introduction}
\label{sec:introduction}

Historical visualizations hold a pivotal role in the study of visualization history. 
By examining the visualization practices and design principles of different historical periods, researchers can gain a deeper understanding of the evolution and development of contemporary visualization methods.
Culture plays a significant role in shaping historical visualizations. 
Different civilizations exhibit distinct features in their visual communication and design aesthetics, which impose distinct requirements on how information is represented and conveyed within each cultural context.
Therefore, studying historical visualizations from diverse cultural perspectives is crucial for transcending singular or biased viewpoints and achieving a more objective and comprehensive understanding.

However, current studies on historical visualizations, whether focused on single case~\cite{Koch2009Crediting, Friendly2002Visions, Friendly2005Early, Wilkinson2009History, Marchese2011Exploring} or large corpora~\cite{Zhang2024OldVisOnline, Friendly2001Milestones}, are limited to eurocentric views~\cite{Gunter2021Review}.
The main reason could be the language barrier and cultural discrepancy~\cite{Friendly2021Discussion}, which hinders the inspection of visualizations across different cultures. 

To bridge this research gap, we investigate historical visualizations within the Chinese cultural context. 
By bringing Chinese historical visualizations into the broader discourse of visualization studies, we seek to stimulate further cross-cultural investigations of historical visualizations and enhance our understanding of diverse visualization practices.

Specifically, we propose \textit{\datasetName}, a collection of historical Chinese visualizations.
It is named after \textit{Zuan Tu} (``\chinese{纂圖}'' in Chinese), the classical Chinese term referring to the list of figures placed at the beginning of a book. 
Historical Chinese books contain diverse visual representations, some conveying data and concepts, while others do not encode data.
In this work, we collect all visual representations and termed them as ``graphics'' with those conveying data classified as visualizations and decorative elements referred to as illustrations.

Our work first collects a corpus of historical Chinese graphics from various sources, including a large number of historical Chinese books. 
In the corpus, there are over \numVisShort visualizations and \numIllusShort illustrations retrieved from data sources with more than \numBookShort books.
Hidden in ancient books and other mediums, these historical Chinese \itemTypePlural cannot be easily searched or accessed online.
The data curation process is introduced in~\cref{sec:preliminary-curation} and~\cref{sec:large-scale-curation}.

Based on the dataset, we study the visual patterns of historical Chinese visualizations and analyze the possible causes behind them in~\cref{sec:zuantuset}.
We envision several usage scenarios for our collection in~\cref{sec:usage-scenarios}. 
We also discuss our experiences working with historical documents under different cultures and share our experiences to guide future endeavors in~\cref{sec:discussion}.

In summary, the contributions of this work are:

\begin{itemize}[leftmargin=3.5mm]
    \item We contribute \datasetName, a large-scale historical Chinese visualization dataset.
        \datasetName is constructed through a semi-automatic pipeline to extract visualizations from historical Chinese books.
        \datasetName can be browsed with a gallery at: \url{https://zuantuset.github.io/gallery}.

    \item We introduce Chinese historical visualization into the field of historical visualization studies, examining its visual characteristics and formative factors. We also discuss the usage scenarios of \datasetName.
      
\end{itemize}

    \section{Related Work}
\label{sec:related-work}

\revised{
This section reviews the literature on historical visualization and datasets of visualization.
}

\subsection{Historical Visualization}

Historical visualizations reflect the societal context of their time, offering rich information for research.
Some prior research focuses on well-known historical visualizations and examines them from \revised{different perspectives}.
Koch~\cite{Koch2009Crediting} discusses the people's attitudes toward John Snow's cholera map~\cite{Snow1855Mode} in 19th-century London.
Shiode~\cite{Shiode2015mortality} \revised{utilizes historical records to quantitatively examine John Snow's waterborne transmission hypotheses.}
By retrospecting ``Napoleon's Grand Army''~\cite{Minard1869Carte}, Friendly reviews the contribution of Charles Joseph Minard and compares modern revisions of this classical visualization~\cite{Friendly2002Visions}.
There is also research on the provenance of different types of visualizations, such as scatter plot~\cite{Friendly2005Early} and heat map~\cite{Wilkinson2009History}. 
\revised{
This line of research focuses on close examination of individual visualizations with a microscopic view.
}

In addition to the microscopic view, some other works take a macroscopic view, reviewing the historical development of visualizations.
Friendly~\cite{Friendly2001Milestones} collects the chronological milestones of historical visualizations to show the development process of visualizations.
Correll and Garrison~\cite{Correll2024When} comprehensively examine the development of historical visualizations and illustrations related to the human body, highlighting the importance of culture in the understanding and reflection of visualizations.
There are also books~\cite{Friendly2008Brief, Friendly2021History, Rendgen2019History} illustrating the development of visualizations.

While these works elaborate on how, where, and why today's data visualization is developed and conceived, they inevitably partially overlook some \revised{``historical devices and trajectories of change that have not directly led to present-day forms''}~\cite{Ruokkeinen2023Developing}.
That is, what is promotive for modern visualization gets more attention and discussion, while earlier attempts that do not contribute much to later statistic diagrams are passed over~\cite{Ruokkeinen2023Developing}.
Our work focuses on historical Chinese visualizations, one of the many overlooked branches.
To this end, we collected a large number of historical Chinese visualizations.
Based on the dataset, we analyze these visualizations (\cref{sec:zuantuset}, \cref{sec:usage-scenarios}, and \cref{sec:discussion}) and sought to disseminate historical Chinese visualizations to a wider audience.
We aim to contribute to addressing the ``concern of eurocentric view''~\cite{Gunter2021Review} in previous work.

\subsection{\revised{Dataset of Visualizations}}

\revised{
With the popularization of data-driven research, datasets of visualizations are created for various purposes.
Specifically, they are useful as corpora for summarizing design patterns and benchmarks for empirical studies.
}
Segel et al. summarize different storytelling techniques from narrative visualization samples~\cite{Segel2010Narrative}.
Zhang et al. collect visualizations related to COVID-19 to discover ``who uses what kinds of data to communicate what messages''~\cite{Zhang2021Mapping}.
Borkin et al. collect \num{2070} visualizations to find out what visualizations are memorable~\cite{Borkin2013What}.
\revised{
Other works collect scholarly visualizations in IEEE VIS and IEEE TVCG papers to summarize common designs and research trends~\cite{Chen2021VIS30K, Deng2023VisImages}.
}
\revised{
Regarding historical visualizations, Friendly et al.'s Milestones Project~\cite{Friendly2001Milestones} gathers hundreds of significant inventions in the history of visualization.
Zhang et al.' OldVisOnline~\cite{Zhang2024OldVisOnline} curates a dataset of 13K historical visualizations.
}

\revised{
Existing historical visualization datasets fall short in that they typically overlook visualizations from non-European cultural frameworks, which may bring bias when using these datasets for analysis~\cite{Friendly2021History}.
This work constructs the first dataset dedicated to historical Chinese visualizations and illustrations as an initial effort to enhance existing historical visualization collections and draw attention to visualizations from underrepresented cultural frameworks.
}
\datasetName includes visualizations \revised{between 550 BCE and 1950 CE} in China.
Through this dataset, we aim to promote historical Chinese visualization and bring this knowledge to broader audiences.

    \section{\revised{Preliminary Data Curation for \datasetName}}
\label{sec:preliminary-curation}

\revised{
Before the large-scale data curation described in \cref{sec:large-scale-curation}, we went through a preliminary data curation process.
Through this process, we manually collected \numPreliminaryRelevantImages images of historical Chinese visual representations from the web and refined our scope for data collection.

\textbf{Manual collection from the web:}
We first manually collected \numChaofanYangRelevantImages images of historical Chinese visual representations from the web.
These visual representations include maps, genealogies, geometry, and paintings.
Their physical forms encompass a variety of materials, including stone, wooden boards, silk, and paper.
Their themes include geography, astronomy, medicine, and genealogy.
As we noticed a manually collected image was from Dunhuang, we looked for more relevant images from Pelliot chinois Dunhuang manuscripts\footnote{Last accessed on Feb 1, 2024 with the search keyword \href{https://gallica.bnf.fr/services/engine/search/sru?operation=searchRetrieve&version=1.2&query=(gallica all "Pelliot chinois")}{``Pelliot chinois''}.} in Gallica~\cite{NLFGallica}, and obtained another \numDunhuangRelevantImages images.
We also manually collected \numShugeBooks historical Chinese books from the Shuge digital library~\cite{Ceng2013Shuge} where we obtained \numShugeRelevantImages images of historical visual representations.

\textbf{Scope:}
Our initial objective of data collection was to gather images corresponding to \taxon{visualization}.
Specifically, we refer to \taxon{visualization} as a visual representation that uses graphical marks to encode abstract or spatial data\footnote{
    In this paper, we adopt a broad notion of \taxon{visualization} that includes visual representations that may not commonly be regarded as \taxon{visualization}.
    For example, despite the controversy on whether \taxon{map} should be categorized as \taxon{visualization}~\cite{Hograefer2020State,Friendly2010First,Chen2024Image,Zhang2025VisTaxa}, we include \taxon{map} as \taxon{visualization}.
    We also include \taxon{table} of structured data as \taxon{visualization}.
}.
Through our data collection practice, we observed that many figures presented in historical Chinese books did not fall into conventional notions of \taxon{visualization}.
These \taxon{non-visualization} figures generally follow the specification of \taxon{illustration} in Zhang et al.'s taxonomy~\cite{Zhang2025VisTaxa} that defines \taxon{illustration} as ``a visual representation that commonly uses drawings, sketches, or paintings to represent visual explanations of concepts, plans, processes, or scenes''.
We decided to include \taxon{illustration} as part of our dataset under two considerations.
Firstly, \taxon{illustration} may also serve some of the usage scenarios described in \cref{sec:usage-scenarios}.
Secondly, the border between \taxon{visualization} and \taxon{illustration} can be subjective, which is further discussed in~\cref{sec:lesson}.
By including \taxon{illustration} into our dataset, we aim to provide a more comprehensive dataset that users may redefine the boundary between \taxon{visualization} and \taxon{illustration} based on their needs.
Throughout the writing, we use \taxon{graphic} to refer to both \taxon{visualization} and \taxon{illustration}.

As we decided to expand the collection to include more data sources, the manual collection process was no longer scalable.
To improve the efficiency, we implement a semi-automatic pipeline for large-scale data curation, as described in \cref{sec:large-scale-curation}.
}

    \begin{figure*}[!ht]
    \centering
    \includegraphics[width=\linewidth]{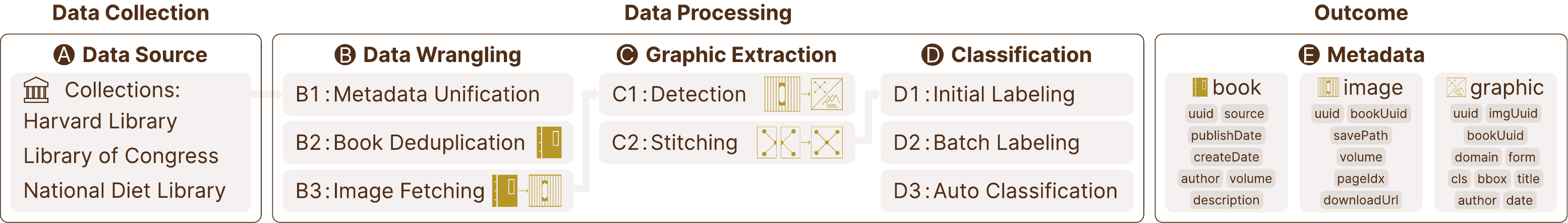}
    \caption{
        \revised{
 The semi-automatic data curation pipeline for historical Chinese visualizations and illustrations:
 (A) We collect historical Chinese books from collections of three digital libraries.
 (B) We unify and deduplicate book metadata from different libraries, then fetch corresponding images.
 (C) We utilize YOLO to detect \itemTypePlural from these images. Some incomplete \itemTypePlural are stitched.
 (D) The \itemTypePlural are further classified through three steps. 
 Initial labeling aims to obtain a basic taxonomy.
 Batching labeling aims to further expand the quantity of labeled \itemTypePlural.
 Similarity-based matching assigns labels to unlabeled \itemTypePlural.
 (E) The data structures for the metadata of three types of entities involved in \datasetName: \schemaBook, \schemaImage, and \schemaGraphic.
 }
        \yzc{
 1. why are some attributes not aligned with OldVisOnline:
 - source -> sources.name, source.url, source.accessDate
 - author -> authors
 2. compile an appendix illustrating the meaning of each data attribute.
 3. we may refactor ``theme'', ``type'', and ``cls'' into a single ``classes''/``categories'' attribute that stores the categories paths
 4. document the exact meaning of each attribute in an appendix
 }
 }
    \Description{
 The pipeline for data curation.
 The pipeline has two main stages: data collection and data processing.
 The data processing stage involves data wrangling, graphics extraction, and classification.
 }
    \label{fig:pipeline}
\end{figure*}

\section{Large-Scale Data Curation for \datasetName}
\label{sec:large-scale-curation}

\revised{
Following the preliminary manual data curation process in \cref{sec:preliminary-curation}, this section introduces a semi-automatic pipeline for large-scale data curation.
\Cref{fig:pipeline} shows the pipeline with two main stages: data collection and data processing.
\Cref{sec:data-source} describes the data sources for data collection (\cref{fig:pipeline}(A)).
The data processing stage involves data wrangling (\cref{fig:pipeline}(B) and \cref{sec:data-wrangling}), graphics extraction (\cref{fig:pipeline}(C) and \cref{sec:graphic-extraction}), and classification (\cref{fig:pipeline}(D) and \cref{sec:classification}).
Through the pipeline, we obtain images of historical Chinese graphics and related metadata (\cref{fig:pipeline}(E)).
Note that the books from Shuge and Dunhuang manuscripts from Gallica collected in the preliminary data curation (\cref{sec:preliminary-curation}) also went through the data processing steps.
\Cref{sec:gallery} introduces the gallery for browsing the resulting \datasetName dataset.
}

\subsection{Data Source}
\label{sec:data-source}

\revised{
To scale our data collection, we focus on the Chinese rare book collections from three digital libraries: \dbname{Library of Congress}~\cite{LibraryCongressLibrary}, \dbname{Harvard Library}~\cite{HarvardLibrary2023Harvard}, and \dbname{National Diet Library}~\cite{NDLRare}.
These data sources provide online APIs for image retrieval.
Different digital libraries have different conventions for storing books.
Some may store a book in multiple collection items.
Thus, for each data source, we go through a custom process to merge items into books.
We obtain \numBooksLoC, \numBooksHL, and \numBooksNDL books from these digital libraries, respectively.
For each book, we obtain their metadata and their corresponding image resources stored as IIIF manifests~\cite{IIIF2011International}.
}

\future{Document the raw data structure from the digital libraries in an appendix (like the appendix of OldVisOnline).}
\future{report the number of duplicates and the number of different editions of the same book}

\subsection{Data Wrangling}
\label{sec:data-wrangling}

\revised{
To merge the data obtained from different sources, we go through metadata unification, book deduplication, and image fetching.

\subsubsection{Metadata Unification}

We normalize the metadata obtained from the data sources into three schemas: \schemaBook, \schemaImage, and \schemaGraphic.
\Cref{fig:pipeline}(E) shows the attributes stored for each schema.

\begin{itemize}[leftmargin=3.5mm]
    \item A \schemaBook instance corresponds to a book from the data sources.
    
    \item An \schemaImage instance corresponds to a bitmap image of one or multiple pages in a book.
 An \schemaImage instance is usually associated with a \schemaBook instance through the \inlinecode{bookUuid} attribute.
 Each \schemaBook instance corresponds to at least one and typically multiple \schemaImage instances.
 Note that the \schemaImage instances obtained from the web and Gallica in \cref{sec:preliminary-curation} are not associated with any \schemaBook instance.
 Some of these images are not from books, and the original books of the others are hard to identify.
    
    \item A \schemaGraphic instance corresponds to a \taxon{visualization} or \taxon{illustration} detected from an \schemaImage instance.
    \Cref{sec:graphic-extraction} details the detection method.
 A \schemaGraphic instance is associated with an \schemaImage instance through the \inlinecode{imgUuid} attribute.
 Each \schemaImage instance may correspond to no, one, or multiple \schemaGraphic instances.
\end{itemize}

The data structure was adapted from the OldVis schema~\cite{Zhang2024OldVisOnline}.
Note that the original OldVis schema only concerns the image entity and does not consider the concepts of books containing images and graphics detected from images.
}

\subsubsection{Book Deduplication}

\revised{
After obtaining the \schemaBook instances, we deduplicate them through a semi-automatic matching process.

Let $len(x)$ be the length of arbitrary string $x$.
Let $lev(x, y)$ be the Levenshtein distance function between arbitrary strings $x$ and $y$.
Let strings $a$ and $b$ be the \inlinecode{title} of two \schemaBook instances.
Let strings $a'$ and $b'$ be the concatenations of \inlinecode{title}, \inlinecode{publishDate}, and the number of associated \schemaImage instances of two \schemaBook instances.
We find all pairs of \schemaBook instances that satisfies both $1 - \frac{lev(a, b)}{len(a) + len(b)} > 0.9$ and $1 - \frac{lev(a', b')}{len(a') + len(b')} > 0.95$.

Then, we manually verify all the resulting pairs of \schemaBook instances, which are potential duplicates.
Note that we do not consider different editions of the same book to be duplicates.

\subsubsection{Image Fetching}

With the stored metadata of \inlinecode{book} and \inlinecode{image}, we fetch the corresponding images from the digital libraries with \inlinecode{downloadUrl} attribute values.
}

\subsection{Graphics Extraction}
\label{sec:graphic-extraction}

\revised{
As described in \cref{sec:preliminary-curation}, we use \taxon{graphic} to refer to both \taxon{visualization} and \taxon{illustration}.
The following introduces the two steps to extract \schemaGraphic instances from \schemaImage instance: detecting graphics from images and stitching parts of graphics split across pages.
}

\subsubsection{Detection}
\label{sec:detection}

We fine-tuned the YOLOv8l model~\cite{Jocher2023Ultralytics} to detect graphics from images.
\revised{
As there is no training data on historical Chinese graphics, we conducted an iterative process of labeling, model fine-tuning, and detection on the remaining images.
}

We manually selected 8 books from Shuge~\cite{Ceng2013Shuge} with abundant visual contents and different themes.
From these books, we manually labeled the bounding boxes and types (\taxon{visualization} or \taxon{illustration}) of graphics with the Roboflow Annotate tool~\cite{RoboflowRoboflow}.
From the books, we obtained 976 images containing graphics.
\future{Whether 976 refers to the number of graphics or the number of images containing graphics needs to be checked.}
These images were then combined with \num{100} randomly sampled images containing no graphics, forming the initial training set with \num{1076} images.
The initial training set was then augmented to \num{2288} images using random cropping and blurring.
We use the initial training set to fine-tune YOLOv8l pre-trained on the COCO dataset~\cite{Lin2014Microsoft}.
\future{We may look for the model performance for the first round.}

The fine-tuned model was then used to detect graphics from \num{400} of the books in \dbname{Library of Congress}~\cite{LibraryCongressLibrary}, identifying \num{13782} potential graphics in total.
We then manually correct the detected result.
All the images containing true positive graphics and \num{700} randomly sampled images containing false positives combined with the initial training set to form the enlarged training set.
The enlarged training set contains \num{6783} images with \num{10056} graphics.
We fine-tuned YOLOv8l on the second dataset, which achieved \recall recall, \precision precision, and \Fone F1 score.
We then applied the fine-tuned models to the remaining unlabeled images from books in \dbname{Library of Congress}~\cite{LibraryCongressLibrary}, \dbname{Harvard Library}~\cite{HarvardLibrary2023Harvard}, and \dbname{National Diet Library}~\cite{NDLRare}.

Note that we do not include the titles in the bounding box of graphics.
The consideration is that when fine-tuning the model, including the titles in the bounding box may lead the model to learn to detect textual patterns.

\subsubsection{Stitching}
\label{sec:stitching}

Among the graphics we obtained, some graphics corresponded to the same visualization but were split on separate pages of a book.
We thus stitch such segments across pages.

We grouped instances of \schemaGraphic distributed at consecutive book pages into pairs.
Each pair was then processed by Gaussian blur to denoise and morphological closing to remove detail and small holes.
Next, we used CLIP ViT-B/32~\cite{Radford2021Learning} to compute embedding and calculate the similarity for each pair.
We manually checked the pairs with similarity above 0.95 on whether they should be stitched and used Adobe Photoshop for stitching.

\subsection{Classification}
\label{sec:classification}

\begin{figure*}[!htbp]
    \centering
    \includegraphics[width=\linewidth]{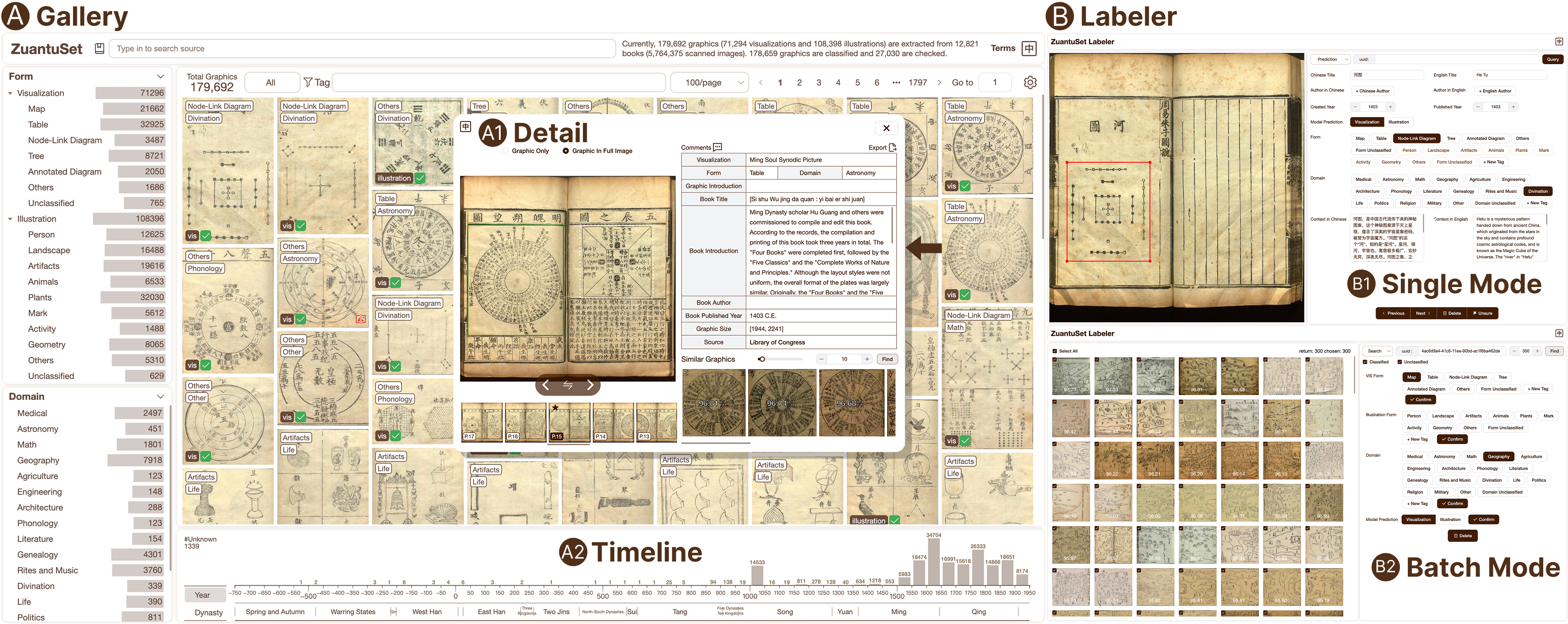}
    \caption{
        \textbf{Two interactive systems of \datasetName:}
        (A) ZuantuSet Gallery for the user to browse historical Chinese graphics.
        \revised{
        (A1) Detail panel of a graphic.
        (A2) Timeline showing the temporal distribution of the filtered graphics.
        (B) ZuantuSet Labeler for the user to categorize historical Chinese graphics.
        (B1) Single mode labeling for the user to edit the bounding box and metadata of a single \itemType.
        (B2) Batch mode labeling for the user to retrieve multiple \itemTypePlural through metadata query or similarity matching and label them at once.
        }
    }
    \label{fig:gallery}
    \Description{The image showcases a user interface for exploring, filtering, and efficiently labeling visualizations based on attributes like source, type, theme, and time.}
\end{figure*}

The object detection model in \cref{sec:detection} classified graphics into two categories: \taxon{visualization} and \taxon{illustration}.
Not every historical Chinese book contains \itemType.
Out of the total \numBook collected books, \numBookHasGraphic books contain \itemType.
As mentioned in \cref{sec:graphic-extraction}, \itemType are classified into \taxon{visualization} and \taxon{illustration}.
Among the books, \numBookHasVis books contain \taxon{visualization}.
If sorting these books with the number of contained visualizations in descending order, the top 84 books contribute around 50\% of the visualizations, and the top 838 books contribute around 90\% of the visualizations.

\revised{
To facilitate retrieval and analysis of the graphics, we aim to classify them into more specific subcategories further.
We perform a three-step classification process: initial labeling, batch labeling, and similarity-based matching.
The first two steps involved manual labeling and are conducted in the \datasetName labeler shown in \cref{fig:gallery}(B).
}
We categorize \taxon{graphic} on two aspects: form and domain.

\begin{itemize}[leftmargin=3.5mm]
    \item \textbf{Form} refers to the visual appearance of the graphic.
 For a \taxon{visualization}, the form corresponds to a chart type.
 For a \taxon{illustration}, the form corresponds to the prominent subject.
    
    \item \textbf{Domain} refers to the application domain of a graphic.
\end{itemize}

\subsubsection{Initial Labeling}

We perform an initial labeling process to develop a preliminary taxonomy of historical Chinese graphics.
In this process, \num{2000} graphics were randomly chosen and labeled by one of the authors using the single mode labeling interface, as shown in~\cref{fig:gallery}(B1).
The interface allows the annotator to create new subcategories for the form and domain aspects.
A leaf node of form and a leaf node of domain can be assigned to each graphic.
The annotator may revise incorrect predictions and add new tags for unseen forms or domains.
In the labeling process, for graphics with unsure labels or belonging to types with few instances, we group them into \taxon{others}.
The resulting taxonomy includes 5 visualization forms, 8 illustration forms, and 16 domains (excluding the \taxon{other} type), detailed in \cref{sec:dataset-sample} and discussed in \cref{sec:lesson}.

\subsubsection{Batch Labeling}

With the developed taxonomy, we used the batch mode labeling interface in~\cref{fig:gallery}(B2) to annotate more images.
The user may retrieve images with a similar visual appearance to a given image.
To identify similar graphics, we use CLIP ViT-B/32~\cite{Radford2021Learning} to compute the graphic embeddings and calculate the cosine similarity between embeddings.
If needed, the user may also search with other query criteria, such as searching for graphics from the same book.
By retrieving images that share similarities under certain criteria, the user may quickly assign labels to multiple graphics simultaneously.
For example, graphics with similar visual appearance may share the same form, and graphics in the same book may share the same domain. 
In total, we manually labeled \numChecked graphics with both form and domain through initial labeling and batching labeling.

\subsubsection{Similarity-Based Matching}

We use CLIP ViT-B/32~\cite{Radford2021Learning} to label forms of the remaining graphics based on similarity.
Each unlabeled \itemType is assigned the same labels as its most similar labeled graphic.
Through the three classification steps, we obtain the labels of \numClassified \itemTypePlural.

\subsection{Gallery}
\label{sec:gallery}

To present the outcome of our data curation process, we provide an online gallery as shown in \cref{fig:gallery}(A).
Users can select \itemTypePlural within the gallery by filtering forms, domains, and sources for visualizations and illustrations.
By clicking a \itemType thumbnail, the detail panel pops up, as shows in~\cref{fig:gallery}(A1).
The timeline (\cref{fig:gallery}(A2)) shows temporal distributions of selected \itemTypePlural. 

    \begin{figure}[!ht]
    \centering
    \includegraphics[width=\linewidth]{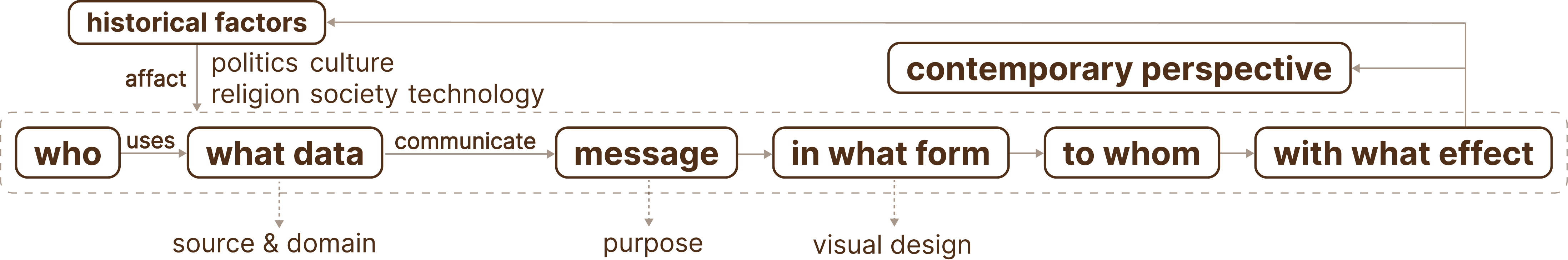}
    \caption{
        \textbf{A framework of elements involved in historical visualizations as a visual communication channel:}
        The framework is adapted from frameworks in the literature~\cite{Zhang2021Mapping, Munzner2009Nested,Lasswell1948Structure}.
        We emphasize the impact of historical factors, such as politics, cultures, and religions, on the framework components.
        \revised{We also consider the influence on contemporary perspective when investigating the effect of historical visualizations.}
    }
    \Description{
        The framework for understanding historical visualizations builds on prior research, focusing on the influence of historical factors, such as politics, culture, and religion.
        It categorizes feedback into its impact on these historical factors and its effects on the current state of study.
    }
    \label{fig:framework}
\end{figure}

\section{\revised{Analyzing \datasetName}}
\label{sec:zuantuset}

\revised{
This section focuses on the content of \taxon{visualization} images in \datasetName.
}
We analyze historical Chinese visualizations based on the framework in \cref{fig:framework}, which is adapted from the literature~\cite{Zhang2021Mapping,Munzner2009Nested,Lasswell1948Structure}.
The framework is devised for \revised{understanding elements involved in historical visualizations as a visual communication channel}.
The primary function of the original framework~\cite{Zhang2021Mapping} is to organize and analyze surveyed visualizations, providing concepts that guide further inquiry and capture their commonalities.
\revised{
We adapt the framework to emphasize historical factors and the effect of historical visualizations on contemporary perspective.
}
\emph{Who} focuses on the creator of \itemTypePlural.
\emph{What data} refers to the domain of the data.
\emph{What message} refers to the desired purpose of \itemTypePlural.
\emph{What form} refers to the visualization taxonomy and visual patterns.

\revised{
\Cref{sec:dataset-sample} discusses the relations between form and domain of historical Chinese visualizations, corresponding to \emph{what data}.
\Cref{sec:taxonomy} focuses on \emph{what message} and \emph{what form} by analyzing visual patterns of historical Chinese visualizations together with historical factors.
}

\subsection{Visualization Forms and Domains}
\label{sec:dataset-sample}

\begin{figure}[!htb]
    \centering
    \includegraphics[width=1.0\linewidth]{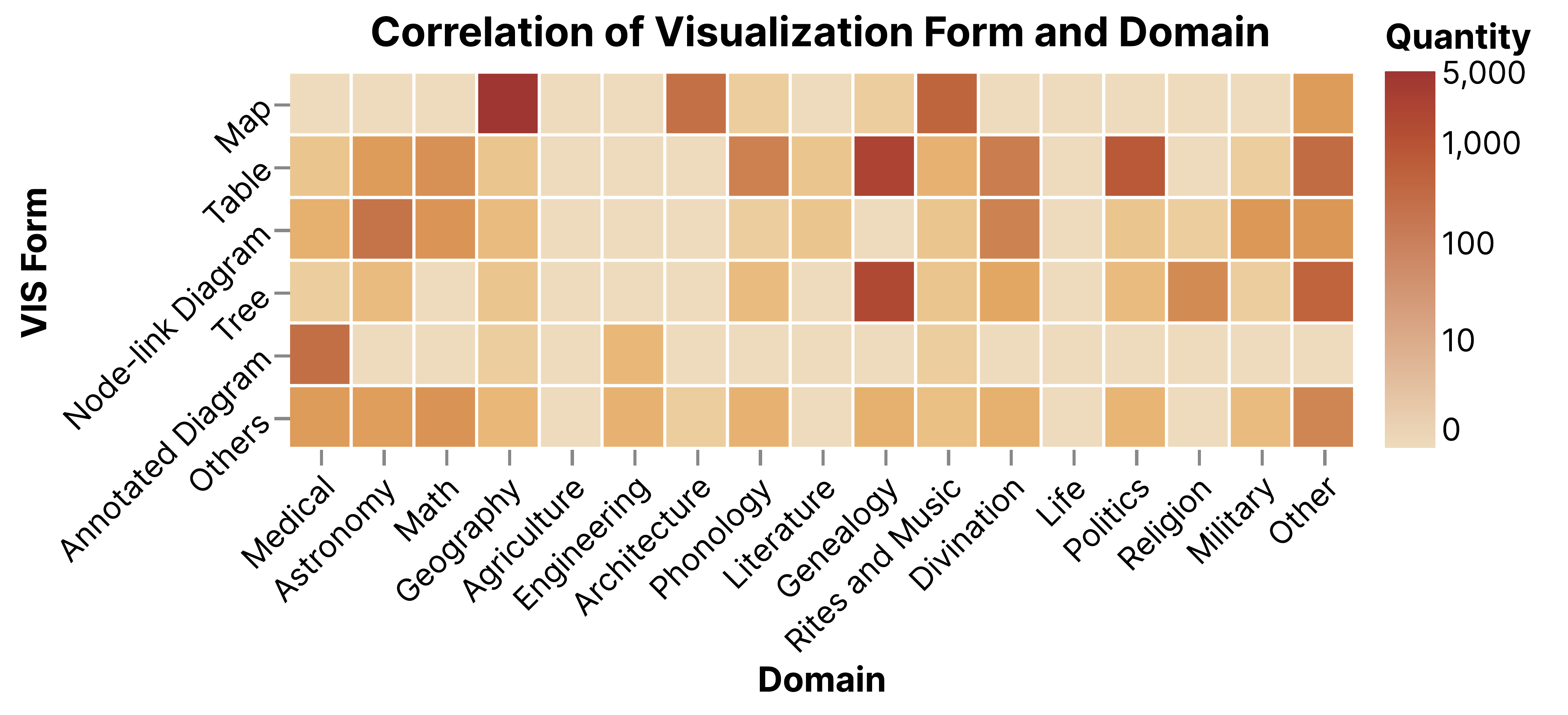}
    \caption{
        The correlation between form and domain of visualizations in \datasetName.
        \yzc{Give the exact number of each cell, each row, and each column.}
    }
    \Description{
        The image describes the connection between visualization forms and domains that can guide collection efforts.
    }
    \label{fig:relation-matrix}
\end{figure}

\Cref{fig:relation-matrix} shows the distribution and correlation of visualization forms and domains in \datasetName.
We observe patterns, such as maps correlating with geography and tables and trees correlating with genealogy.
Through closer inspection, we observe that different types of books contribute visualizations differently in forms and domains.

\textbf{Books focusing on one form and domain:}
Some books focus on visualizations of a specific form and domain.
For example, visualizations in \ChineseTerm{\GuangYuJi}{廣輿記}{Enlarged Terrestrial Records}\footnote{
We use the template ``\ChineseTerm{Chinese Phonetic Alphabet}{Chinese term}{English translation}'' to introduce Chinese book and figure titles and terms.
}~\cite{Lu16621722Guang}, a Chinese geography book, are all maps.
\ChineseTerm{\DongTingQinShiZongPu}{洞庭秦氏宗譜}{The Genealogy of the Qin Family in Dongting}~\cite{Qin1873Dongting} recording family genealogy contribute thousands of family trees and tables, and \ChineseTerm{\BenCaoGangMu}{本草綱目}{The Compendium of Materia Medica}~\cite{Li1655Ben}, a famous medical book in ancient China, contributes thousands of illustrations of plants and animals but contains almost no visualizations.

\textbf{Books spanning multiple forms and domains:}
Other books, especially \ChineseTerm{\LeiShu}{類書}{a kind of reference book consisting of material quoted from many sources and arranged by category}, such as \ChineseTerm{\SiShuTuKao}{四書圖考}{Diagram Collection of the Four Books}~\cite{Du1827Si} and \ChineseTerm{\SiShuWuJingDaQuan}{四書五經大全}{Corpus of the Four Books and the Five Classics}~\cite{Hu1403Si} contribute visualizations across a wide range of domains.

\textbf{Where to find certain types of graphics:}
The correlation (\cref{fig:relation-matrix}) between form and domain suggests that to look for a specific form of visualization, and we may look into the book belonging to a highly correlated domain.
For example, genealogy books are the best source for looking for more tables and trees.

\subsection{Historical Chinese Visualizations}
\label{sec:taxonomy}

This section discusses the five different \revised{forms} of historical Chinese visualizations categorized through the classification process in \cref{sec:classification}: \taxon{map}, \taxon{node-link diagram}, \taxon{tree}, \taxon{table}, and \taxon{annotated diagram}.
For each form, we clarify its definition concerned in this paper.
We then give examples and summarize the characteristics of each form and its historical background through literature reviews.

\begin{figure*}[!htb]
    \includegraphics[width=\linewidth]{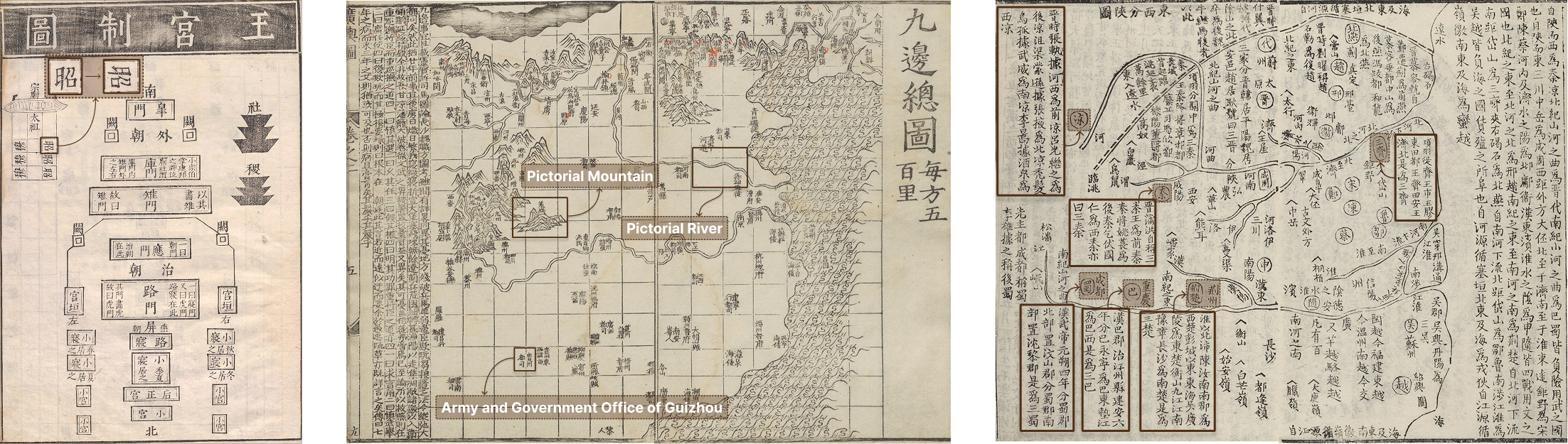}
    \caption{
        \textbf{Examples of map in \datasetName:}
        (Left)
        \ChineseTermWithUrl{\WangGongZhiTu}{王宮制圖}{Layout Plan of the Royal Palace}{https://www.digital.archives.go.jp/acv/auto_conversion/conv/jp2jpeg?ID=M2021050616304107841&p=6} from \term{Qi jing tu}~\cite{Wu1615Qi}, 1615.
        The layout plan shows a planimetric mode that rectangle 
        In the top-left corner is an ancestral temple dedicated to the spirits of deceased ancestors. 
        The annotated Chinese character may represent a tablet bearing the ancestors' names, which is rotated 90 degrees to indicate that the tablet may face the central point.
        (Middle)
        \ChineseTermWithUrl{\JiuBianZongTu}{九邊總圖}{Map of the Nine Garrisons}{https://ids.lib.harvard.edu/ids/iiif/23518593/full/full/0/default.jpg},
        from \term{Guang yu tu}~\cite{Zhu1566Guang}, 1566.
        The map shows the system built during the Ming dynasty (1368 - 1644) to protect the northern border and the Great Wall.
        We annotate pictorial elements such as mountains and rivers.
        Geographic location can be encoded by the scattered texts alone.
        The position of the annotated text indicates the location of the army and government office of Guizhou.
        The map is presented with grids whose length corresponds to 500 Li \revised{(A traditional Chinese unit of distance)}.
        (Right)
        \ChineseTermWithUrl{\DongXiFenShanTu}{東西分陜圖}{Map of Shaanxi}{\UrlOfDongXiFenShanTu}, from \term{Tian xia shan he liang jie kao}~\cite{Xu1723Tian}, 1723.
        Paragraphs are directly written on the map, which provides additional historical information about the specific location.
        We add annotations to the map to indicate some of these paragraphs and point them to the corresponding locations.
    }
    \Description{
        Examples of maps in \datasetName.
    }
    \label{fig:map}
\end{figure*}

\subsubsection{Map}

Map communicates geographical or location information.
In \datasetName, we found a variety of maps, such as geographical maps and layout plans.
We contrast map designs on a spectrum from quantitative to qualitative.

\textbf{Quantitative map design:}
Quantitative design is characterized by the use of  ``scientific'' measurements.
Such maps are typically used for utilitarian purposes, such as administration and military planning~\cite{Yee1994Reinterpreting}.
The quantitative design is common in modern maps that use cartographic approaches, such as the Mercator projection, and the adoption of more standardized cartographic norms~\cite{Yee1994Reinterpreting, Amelung2007New, Yee1994Traditional}.
Quantitative design was uncommon in Chinese maps until the exposure to Western maps in the late modern period when historical governors were aware of the power of the maps' practical function, and the need for more accurate maps increased~\cite{Amelung2007New}.

\textbf{Qualitative map design:}
Before modernization, most historical Chinese maps were characterized by qualitative design, which features textualism~\cite{Yee1994Taking, DorofeevaLichtmann2004Spatial} and pictorialism~\cite{Osawa2016Landscape, Yee1994Reinterpreting, Jiang2017heritage}.
Textualism refers to both a reliance on texts as sources of information in the compiling of maps and a reliance on text to complement the presentation of information in maps~\cite{Yee1994Taking}.
Pictorialism refers to the wide use of pictorial elements in maps~\cite{Osawa2016Landscape, Yee1994Reinterpreting, Jiang2017heritage}, from minor decorations to main information carriers.
The tendencies of textualism and pictorialism are also seen in other forms of historical Chinese visualizations.
In the following, we discuss four features of historical Chinese maps related to qualitative design.

\begin{itemize}[leftmargin=3.5mm]
    \item \textbf{Text accompaniment:}
          In historical Chinese cartography, text might not be merely an auxiliary element but a core component of the cartographic representation.
          According to the targeted task, maps may use texts to report corresponding local conditions~\cite{Yee1994Reinterpreting}.
          For instance, a map that is produced to report current local development may consist of texts indicating population, revenue, and ranking~\cite{Osawa2016Landscape}.
          Furthermore, text alone can form the main component of a map expressing geographic information (e.g., \cref{fig:map} (Right)).
          In this case, the Chinese characters may encode positional, directional, and semantic information.

          \item \textbf{Pictorial and planimetric \revised{design}:}
          Most historical Chinese maps were descriptive and used a lot of pictorial elements (styles of depicting the real world).
          Opposite to pictorial is planimetric, which uses a certain level of abstraction to characterize the geometry of real objects.
          In different maps, the ratio between these two styles varies.
          Typically, the planimetric approach was used to characterize roads and rivers (e.g., by single or double line \cref{fig:map}).
          The pictorial element is commonly used for buildings, landscapes, and mountains, serving as decorations.
          In local gazetteers, many landscape-style maps pictorially depict the local environments and affairs with a limited number of texts naming the places or objects. \yzc{Given reference or example to support this argument.}

    \item \textbf{\revised{Coordinate system}:}
    Before the introduction of Western cartography, most historical Chinese maps lacked scale and coordinate grids. 
    Although the ancient Chinese cartographer Pei Xiu introduced grids and \ChineseTerm{Fen Lu}{分率}{The Graduated Divisions}~\cite{needham1974science, Yee1994Reinterpreting}, the prototype of scale in China, and some maps used grids (\cref{fig:map} (Middle)), the scale of these grids were often inconsistent in a map~\cite{Yee1994Reinterpreting} and did not correspond to the meridian and latitude coordinate systems~\cite{Yee1994Reinterpreting}. 
    A possible explanation of this characteristic is that ancient Chinese maps did not primarily focus on geographical accuracy but rather on serving political, religious, artistic, or utilitarian purposes~\cite{Yee1994Reinterpreting}. 
    For example, a map can be used to demonstrate China's territorial integrity and the extent of its administrative authority~\cite{Yee1994Reinterpreting}, focusing on conveying political information rather than precise geographical measurements. 
    In this case, a scale or grid is not required.
    Furthermore, since ancient Chinese maps were often combined with textual annotations, the accompanying text provided quantitative information, which may mean that maps did not rely heavily on scales or coordinate grids to convey quantitative distance data~\cite{Yee1994Reinterpreting}.

    \item \textbf{Orientation:}
          Historical Chinese maps had a unique way of representing orientations.
 Unlike the current convention that the top indicates north, there was no standard for orientation in historical Chinese maps~\cite{Baur2019Cultural}.
 In these maps, the north may be pointed downward or even leftward.
 To indicate directions, historical Chinese maps rarely used arrows but a special way of rotating visual elements or texts within the map~\cite{Baur2019Cultural}.
 That is, the direction of a person or building is represented by the direction of the character or the rotated angle of the building.
 The premise of the practice is the multidirectionalism of Chinese characters that rotated characters are still recognizable and with distinct borders with each others~\cite{Baur2019Cultural}.
          \yzc{What is ``multidirectionalism''?}
          \yzc{characters in other languages are also recognizable after rotation. What is special for Chinese characters?}
 This special representation can be observed in many layout plans (\cref{fig:map}) that show the structure of a royal palace or ritual affairs in daily life.
\end{itemize}

The causes of the above characteristics vary. \yzc{This sentence does not add value.}
In ancient China, the primary producers of maps were the government and elites~\cite{Osawa2016Landscape}.
The lack of numerical measurement may rest with the government purpose, whose priority is neither the representation of precise nature and reality~\cite{Yee1994Taking} nor the curiosity to explore and describe the unknown lands~\cite{Osawa2016Landscape}, but the perpetuation of political power~\cite{Yee1994Taking}.
The primary function of historical Chinese maps was to report local conditions and customs, economic development, and records of resource exploration to the central government~\cite{Osawa2016Landscape}.
Also, as local officers had widely used the pictorial representation as a tradition, central governments collected and reprinted them further to disseminate the usage of pictorial elements~\cite{Osawa2016Landscape}.
The above features, while being criticized as ``immature and backward'' by some scholars~\cite{Yee1994Reinterpreting}, show that historical Chinese maps should be investigated under a more diverse opinion where the excellence of cartographic is measured by not only scientific purposes but also social, aesthetic, and even religious standards~\cite{Yee1994Reinterpreting, Jiang2017heritage}.

\begin{figure}[!htb]
    \centering
    \includegraphics[width=\linewidth]{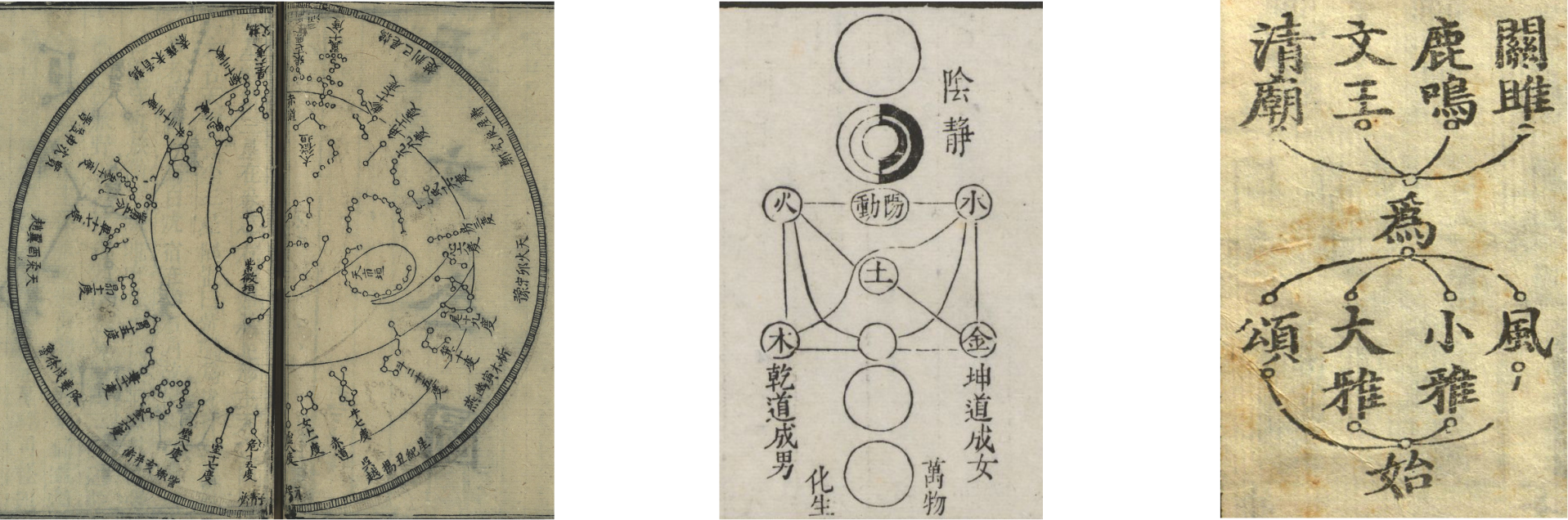}
    \caption{
        \textbf{Examples of node-link diagram in \datasetName:}
        (Left)
        A
        \href{https://tile.loc.gov/image-services/iiif/service:asian:lcnclscd:2011457019:1A002:002b003a/full/pct:100/0/default.jpg}{star chart}
        from \term{Yuechi Xianzhi}~\cite{Xiong1850Yue}, 1850.
        Circles corresponding to stars are connected to form zodiac signs.
        (Middle)
        \ChineseTermWithUrl{Yiyou Taijitu}{易有太極圖}{\YiYouTaiJiTu}{https://ids.lib.harvard.edu/ids/iiif/17209716/full/full/0/default.jpg}, 
        from \term{Lui jing tu kao}~\cite{Yang1662Liu}. 
        It was devised by Zhou Dunyi, a Song dynasty philosopher, synthesizing aspects of Chinese Buddhism and Taoism with metaphysical discussions~\cite{Adler1999Zhou}.
        (Right)
        \ChineseTermWithUrl{\SiShiTu}{四始圖}{The Four Starts of \ShiJing}{https://tile.loc.gov/image-services/iiif/service:asian:lcnclscd:00510373:1D000:34b35a/full/pct:100/0/default.jpg}
        from \term{\SiShuWuJingDaQuan}~\cite{Hu1403Si}, 1403.
        \ChineseTerm{\ShiJing}{詩經}{Classic of Poetry} has four sections.
        \term{\SiShiTu} connects the name of each section and the first poem of each section, forming a knowledge network.
    }
    \Description{
        Examples of node-link diagrams in \datasetName.
    }
    \label{fig:node-link}
\end{figure}

\subsubsection{Node-link Diagram}

In this paper, we consider a node-link diagram as a visual representation that uses lines to connect objects and represent implicit relationships between them.
Among the node-link diagram in \datasetName, nodes are frequently presented as text.
The nodes may also be presented as circles, as in conventional design nowadays.

\textbf{Node as text:}
Linked texts can present textual knowledge.
According to Lackner~\cite{Lackner2007Diagrams}, in such designs, key terms are frequently placed in the center with lines guiding the reader in various directions to other textual elements. 
For example, \cref{fig:node-link} (Right) visualizes the relationship of the four sections of \ChineseTerm{\ShiJing}{詩經}{Classic of Poetry} and the first poem of each section.
The four sections from right to left in the figure are \chinese{風}, \chinese{小雅}, \chinese{大雅}, and \chinese{頌}.
The corresponding first poems are \chinese{關雎}, \chinese{鹿鳴}, \chinese{文王}, \chinese{清廟}.
In \cref{fig:node-link} (Right), the four initial poems are placed above the center character, ``\chinese{為}'', that functions as ``is'' in English.
The four corresponding first poems are placed below the center character.
The figure should be read following a top-to-bottom vertical order.
For example, the chain \chinese{關雎} $\rightarrow$ \chinese{為} (is) $\rightarrow$ \chinese{風} $\rightarrow$\chinese{始} (start) in \cref{fig:node-link} (Right) means ``\chinese{關雎} is the start of \chinese{風}''.
One of the patterns is that, in such a text-link diagram, terms that repeatedly occur in several parallel textual segments (e.g., \chinese{為}) are placed at the central of the graphic with other terms surrounding~\cite{Lackner2007Diagrams}.
The main objective of this pattern may be to reveal the parallelism among textual segments or to emphasize the crucial message of textual segments lies with the center through demonstrating the network-like structure~\cite{Lackner2007Diagrams}.
The text-link diagram may be used primarily for pedagogical purposes by showing the ``general meaning'' and characterizing a passage as a mnemonic element to help readers quickly remember paragraph~\cite{Lackner2007Diagrams}.

\textbf{Comparison with the contemporary visual convention:}
The design of such historical text-link diagrams looks similar to the SentenTree design~\cite{Hu2017Visualizing}, which is also a node-link diagram with text being nodes and links indicating word co-occurrence.
However, we note that the historical text-link diagrams exhibit a different visual convention from contemporary node-link diagrams.
In contemporary node-link diagrams, when there visually exists a path between two nodes, it indicates there exists a path in the underlying graph data structure.
In contrast, for historical text-link diagrams, such an indication is not valid.
For these diagram, only when there exist a visual path between two nodes, and that the two nodes are vertically aligned, can we infer the existence of a path in the underlying data structure.

\textbf{Node as circle:}
Linked circles can be found in historical Chinese star charts (\cref{fig:node-link} (Left)).
Although star charts carry positional information, their visual representation highly resembles a conventional node-link diagram, and the linking conveys the conceived connection among stars.
Others may utilize node-link diagrams to present the relationship of concepts in Taoist philosophy (\cref{fig:node-link} (Middle)).
With the incorporation of node-link diagrams and Taoist philosophical concepts, Taoist visualizations may historically serve as esoteric materials accessible to practitioners and provide visual aids to facilitate adepts' asceticism~\cite{Huang2015Picturing}.

\begin{figure}[!tb]
    \centering
    \includegraphics[width=\linewidth]{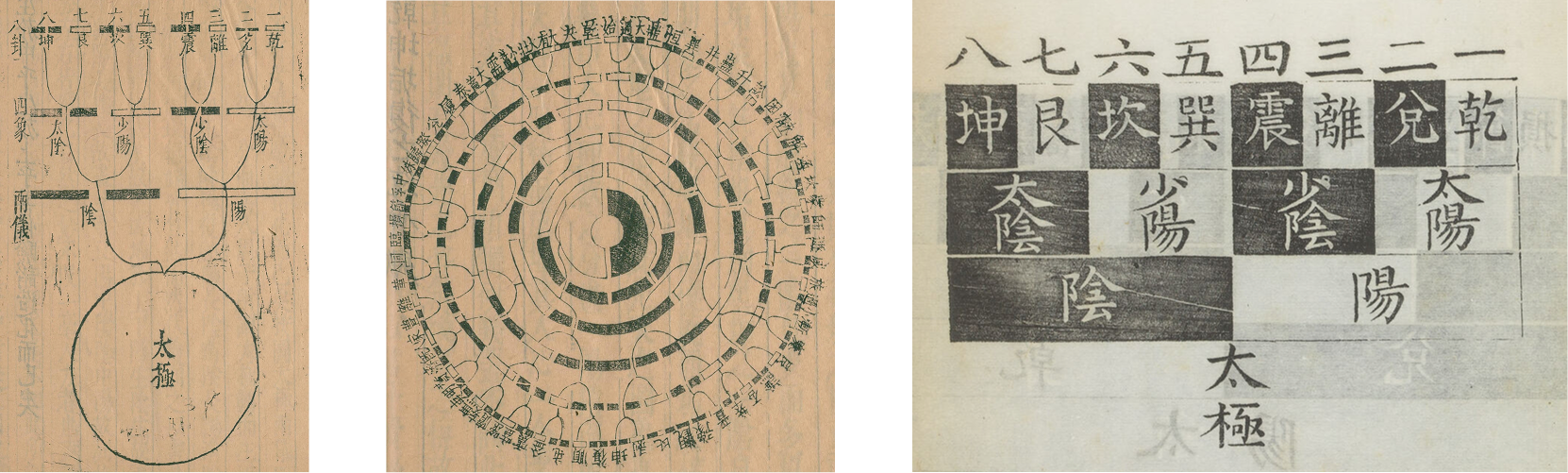}
    \caption{
        \textbf{Examples of tree in \datasetName:}
        The trees visualize the concept of \term{Changes} in \term{I Ching}.
        (Left)
        \href{https://dl.ndl.go.jp/pid/2596441/1/46}{A tree}
        from \term{Tu shu bian}~\cite{Huang1613Tu}, 1613.
        (Middle)
        \href{https://dl.ndl.go.jp/pid/2596441/1/58}{A radial tree}
        from \term{Tu shu bian}.
        (Right)
        \href{https://iiif.lib.harvard.edu/manifests/view/drs:53063658$144i}{An icicle tree}
        from \term{Yu zuan Xing li jing yi}~\cite{Li1717Yu}, 1717.
        \future{Reproduce these figures with their corresponding contemporary design to make their visual encoding clear.}
        \yzc{Instead of using ``Top'' and ``Bottom'' to refer to subfigures, I would suggest adding labels.}
 }
    \Description{
        Examples of trees in \datasetName.
    }
    \label{fig:tree}
\end{figure}

\subsubsection{Tree}

Tree visualizations show hierarchical data.
While hierarchical data may be visualized with a node-link diagram, making tree and node-link diagram potentially overlapping, in this work, we differentiate trees from node-link diagrams.
Our consideration is that \datasetName contains numerous tree visualizations that exhibit characteristics distinct from node-link diagrams.
Historical Chinese tree visualizations mainly consisted of tree visualizations for genealogy and abstract concepts.

\textbf{Visualizing genealogy:}
Genealogy in historical Chinese visualizations covers various domains, including not only family pedigrees but also apprenticeships and knowledge transmission, such as the visualization of apprenticeship relation in \cref{fig:textual-criticism}.

\textbf{Visualizing abstract concepts:}
A tree is also frequently used to visualize abstract concepts in historical Chinese visualizations, such as the concept of \term{Changes} in \ChineseTerm{I Ching}{易經}{Book of Changes}, a divination text in ancient China.
It corresponds to a binary tree structure with the first level representing \ChineseTerm{Taiji}{太極}{Supreme Ultimate}.
\emph{Taiji} generates \ChineseTerm{Liangyi}{兩儀}{Two Modes} on the second level.
The \term{two modes} generate \ChineseTerm{Sixiang}{四象}{Four Images} on the third level.
The \term{four images} generates \ChineseTerm{Bagua}{八卦}{Eight Trigrams} on the fourth level.
This binary tree may also be extended to the seventh level, resulting in 64 \ChineseTerm{Gua}{卦}{Hexagrams}.
The hierarchy can be visualized as trees with different designs, as \cref{fig:tree} shows.
This visualization can also be found with the accompaniment of other iconic concepts (e.g., the \term{Celestial Stems}, the \term{Sexagenary Cycle}, and the \term{Five Phases}) in China~\cite{Kalinowski2007Time}.
The functions of these divination diagrams vary from fortune-telling to the origin of binary system~\cite{Cammann1991Chinese}, and the actual use of them remains in debate.

\begin{figure*}[!htb]
    \centering
    \includegraphics[width=\linewidth]{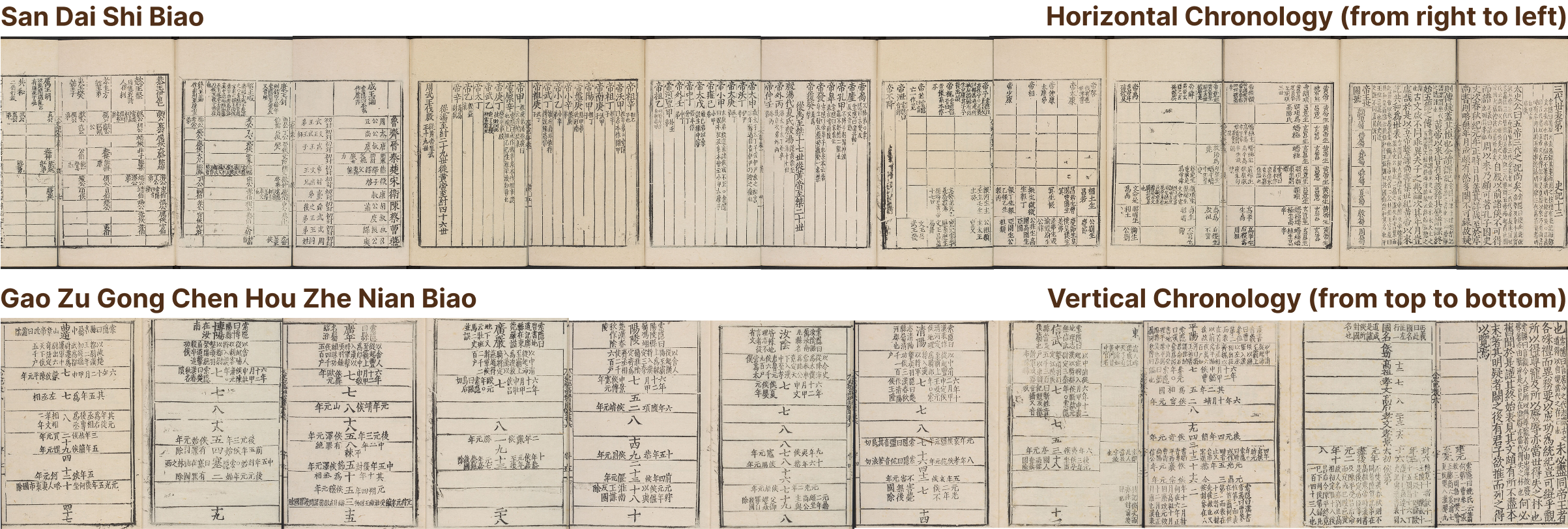}
    \caption{
        \textbf{Examples of tables in \datasetName:}
        The chronological tables are from  \term{\Shiji}~\cite{Si1550Shi}
        (Top) \ChineseTermWithUrl{\SanDaiShiBiao}{三代世表}{Genealogical Table of the Three Ages}{https://iiif.lib.harvard.edu/manifests/view/drs:19140085408i} is an example of horizontal chronological table.
        The chronological order is from right to left.
        Note that this order is consistent with the right-to-left writing system of ancient Chinese.
        (Bottom) \ChineseTermWithUrl{\GaoZuGongChenHouZheNianBiao}{高祖功臣侯者年表}{Yearly Table of the Officials who became Marquises in the Time of Gaozu}{\UrlOfGaoZuGongChenHouZheNianBiao} is an example of a vertical chronological table.
        The chronological order is from top to down.
        (Due to space limit, the figure is incomplete.)
    }
    \Description{
       Examples of tables in \datasetName.
    }
    \label{fig:tables-in-shiji}
\end{figure*}

\subsubsection{Table}

In this work, we regard tables as visualization following the view~\cite{Riggsby2019Mosaics}, which suggests that the meaning of tables lies in their matrix-like structure rather than the data points within the cell.
In the following discussion, we show that historical Chinese tables are very much in line with this view in that their structures vary and serve different purposes.

Overall, we observe three types of historical Chinese tables in \datasetName: vertical, horizontal, and radial.
The vertical and horizontal tables were widely used in books related to history and social statistics, such as \ChineseTerm{\Shiji}{史記}{Records of the Grand Historian} and local gazetteers.
Due to the traditional Chinese book production methods, these long tables were usually divided into several pages.
Tables in \term{\Shiji} were usually employed to record administrative affairs such as enfeoffments and appointments~\cite{Vankeerberghen2007Tables}.
It has two forms of tables, which indicate different chronological order.
For the horizontal table (\cref{fig:tables-in-shiji} (Top)), readers should follow a temporal order from right to left (traditional Chinese reading direction) to read the chronology of a country or a dynasty, with the cells listed from the very right-hand side of the document rows showing the names of the country or emperors.
In this case, the horizontal table functions as a contemporary horizontal timeline~\cite{Rosenberg2010Cartographies}.
For vertical tables (\cref{fig:tables-in-shiji} (Bottom)), the very right side of the document shows emperors' reigns in chronological order from top to bottom, and each column represents a noble house from right to left~\cite{Vankeerberghen2007Tables}.

While these tables may seem simple to read and compile, their meaning goes beyond the data in the cells in historical contexts.
First, these tables can reveal historical insights. 
For example, in the vertical chronology, compared with a period where the entire column is filled with names of kingdoms, the period with few kingdoms may imply a dramatic loss of both territory and autonomy~\cite{Vankeerberghen2007Tables}.
This comparison is similar to grids with different color opacity in a modern heat map.
Second, political considerations may influence the choice of vertical and horizontal tables.
A horizontal table was used to present kingdoms because the long period available horizontally helps to present the continuity of the royal family\cite{Vankeerberghen2007Tables}.
The noble houses, however, were placed on vertical tables.
This may be because their fate was determined by the transition of central power (from top to bottom); these houses reflected the bureaucratic hierarchy, and their family continuity was not the main focus~\cite{Vankeerberghen2007Tables}.

Besides \term{\Shiji}, the vertical tables can also be found in family genealogies, where from top to bottom is the family pedigree.
There are also radial tables which can be viewed as the result of bending a horizontal table until the left and right sides are connected.
The original horizontal order (e.g., temporal) was then represented by evenly divided radians.
These tables usually appear in domains related to Chinese traditional solar terms and calendars.

\begin{figure}[!htb]
    \centering
    \includegraphics[width=.3\linewidth]{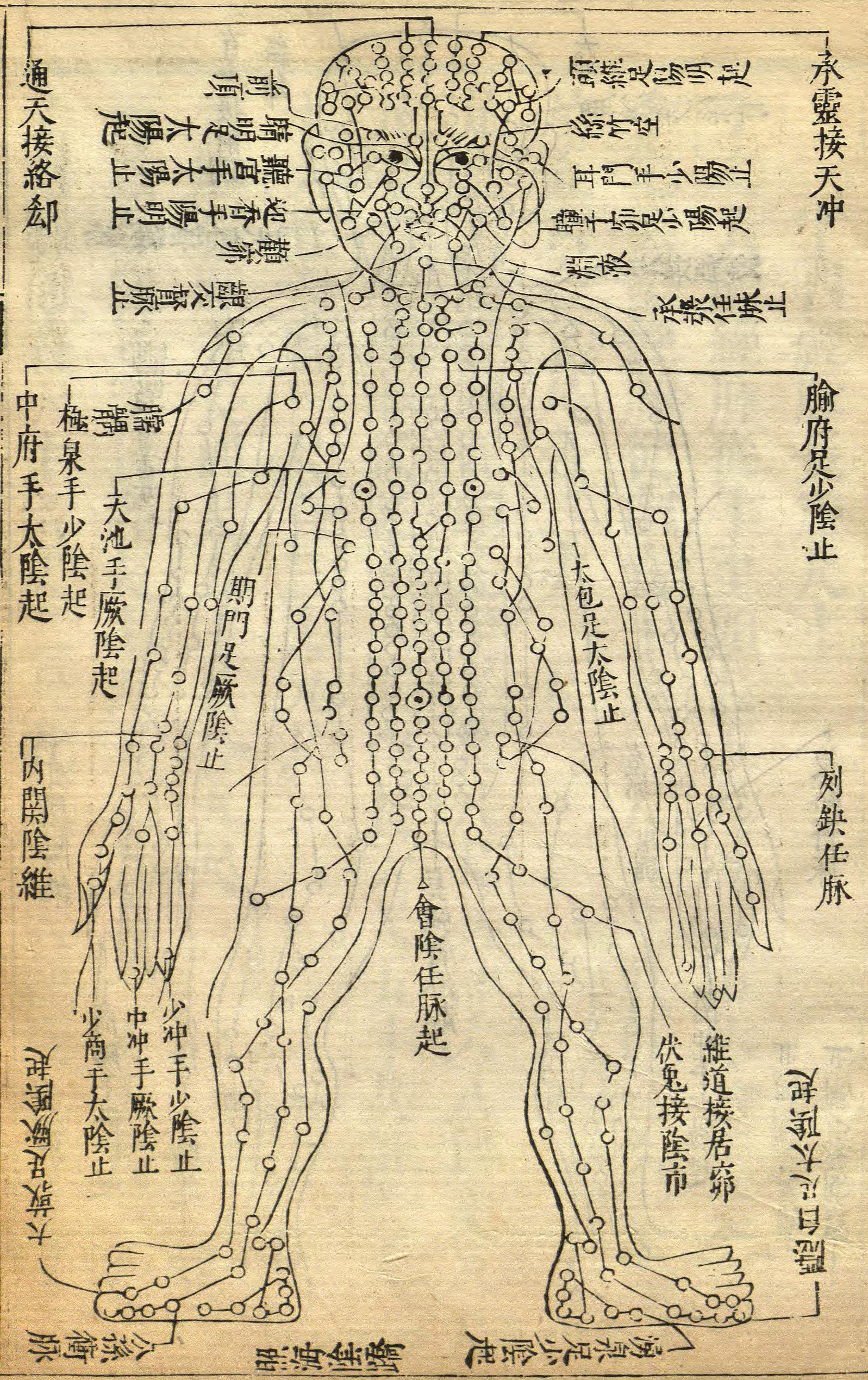}
    \caption{
        \textbf{\href{https://tile.loc.gov/image-services/iiif/service:asian:lcnclscd:2012402918:1A007:01b02a/full/pct:100/0/default.jpg}{An example of annotated diagrams in \datasetName}:}
        From \ChineseTerm{\ZhenJiuDaCheng}{鍼灸大成}{Compendium of Acupuncture and Moxibustion}~\cite{Yang1680Zhen}, 1680.
        In traditional Chinese medicine, different body elements are considered as a whole~\cite{Despeux2007Body}.
        In this visualization, circles representing acupuncture points are connected to form Meridian and Collateral (the passages transporting qi and blood).
        \yzc{This figure is too thin. I would recommend moving part of the caption into the figure as annotation.}
    }
    \Description{
        An example of an annotated diagram in \datasetName.
    }
    \label{fig:illus-diagram}
\end{figure}

\subsubsection{Annotated Diagram}

In this work, we regard annotated diagrams as a special type of illustrations accompanied by labels and lines connecting a concept to a part of the illustration. 
Most historical Chinese annotated diagrams visualize disease and the human body in relation to traditional philosophy.
The contributor to these visuals includes Taoism, divination through body examination, traditional medicine, and forensic medicine~\cite{Despeux2007Body}.
In general, the early Chinese annotated diagrams for the human body focus on depicting the body as a whole~\cite{Huang2015Picturing, Correll2024When} or as a microcosm in the image of the macrocosm~\cite{Despeux2007Body}, while these diagrams in the West pay more attention on musculoskeletal anatomy but lack of the focus on whole systems and features.
This tendency in ancient China can be partially explained by the influence of Taoist philosophy and traditional Chinese medicine.
For Taoism to depict a body, they preferred to emphasize the balance of \term{Yin} and \term{Yang} or the flow of \term{Qi}, and connect spiritual metaphor to the inner world of body~\cite{Despeux2007Body}.
In Chinese medicine, the different elements of the body were considered in relation to each other and within systems of correlation, where they focus on viscera and the circulation of the humours and energies along the meridians~\cite{Despeux2007Body}.
\Cref{fig:illus-diagram} shows the acupuncture points as well as Meridian and Collateral of a whole human body, which reflects concepts in both traditional Chinese medicine and Taoism.
Here, anatomy was not an important part, thus giving rise to the absence of musculoskeletal visuals.
Hence, traditional Taoism and traditional Chinese medicine are crucial factors that influenced the styles of historical Chinese annotated diagrams.

\subsection{Reflections}

\textbf{On overall patterns:}
Ancient Chinese can be regarded as a large discourse community that has its visual conventions~\cite{Kostelnick2003Shaping}.
The visual conventions can be reflected by \revised{the widespread understanding from ancient Chinese people on} visual coding methods and the preference for using pictorial representations in visualizations.
\Cref{sec:taxonomy} discusses that historical Chinese visualizations might serve as political reports, mnemonic elements, and pedagogical resources, but rarely as an analysis approach.
We observe that these visualizations prioritized visual communication over visual analysis.
Similarly, the data visualized in ancient China tended to be imaginary, conceptual, and relational rather than numerical.
Additionally, many historical Chinese visualizations adhere to textualism and pictorialism, which are characterized by the extensive use of pictorial elements in maps and annotated diagrams, non-linear texts in maps (where texts serve as glyphs), trees (where texts serve as leaves of a tree), node-link diagrams (where texts serve as nodes), and annotated diagrams (where texts serve as labels).

\textbf{\revised{On the framework}:} 
In the \cref{sec:taxonomy}, we primarily discuss historical factors, visualization producers, data, and communicated messages. 
This paper has not discussed the components to whom historical Chinese visualization was read and the effect on them. 
We leave these potential discussions to future researchers, which require a deeper historical background.

The following summarizes the discussions in \cref{sec:taxonomy} according to the framework components.
Regarding who produced these visualizations, many historical Chinese visualizations such as maps, node-link diagrams, and annotated diagrams might have been primarily produced by political groups or social elite classes in ancient China.
In terms of data domain and types, although we observe visualizations covering multiple topics—indicating that visualization had been widely used in ancient China—they were more focused on conceptual relationships rather than quantitative measurements.
Much of the data for these visualizations stems from traditional Chinese classics, such as the \term{I Ching} and the traditional Chinese cosmology constructed the forms and connotations of the \term{Taiji} diagram, and literary classics, such as the \term{\ShiJing}, constructed a Chinese early knowledge network.
These ancient Chinese philosophical ideas shaped, to some extent, the emphasis of visualizations on symbolic and qualitative relationships.
We aim to highlight the importance of incorporating historical and cultural contexts into the process of visualization creation and understanding.
We believe that historical factors are essential for understanding historical visualizations under specific ``data cultures'' regarding their formation, functionality, and unique patterns with underlying causes.

\textbf{On comparison with \revised{contemporary} visualizations:}
Here, we summarize the differences between visualizations in ancient China and contemporary according to Baur and Felsing~\cite{Baur2020Visual}.

\begin{itemize}[leftmargin=3.5mm]
    \item \textbf{Visual aesthetics:} Historical Chinese visualizations often prioritize harmonious layouts, flowing compositions, and the integration of textual and pictorial elements, reflecting cultural values of balance and unity~\cite{Baur2020Visual}. 
    In contrast, contemporary visual traditions sometimes emphasize geometric precision, symmetry, and separation of textual and visual elements.
    \yzc{Isn't ``symmetry'' part of ``harmonious layouts''?}
    
    \item \textbf{Purpose:} In ancient China, visualization was more inclined toward the simple display of data, such as in agriculture, handicrafts, and religion, and was influenced by the government~\cite{Yee1994Reinterpreting} and traditional philosophy. 
    Contemporary visualizations are more often used for data analysis, measurements, and exploration, driven by navigation and scientific research~\cite{Baur2020Visual}.

    \item \textbf{Flexibility of visual elements:} Historical Chinese visualizations often emphasized the interdependence within the visual system, with the semantics of visual elements being relatively fixed and less flexible for arbitrary combinations~\cite{Baur2020Visual}. 
    In contrast, contemporary visualizations' visual elements are semantically flexible and capable of conveying multiple meanings through legends and variables such as color and size, making them more suitable for recombination.
    Fixed semantics in ancient China, on the other hand, can evoke more direct cultural and emotional resonance, which is further discussed in \cref{sec:culture-focused-design}.
\end{itemize}

By comparing culture-specific historical visualizations with contemporary designs, we attempt to showcase the importance of constructing a more diverse visualization evaluation system . 
Traditional visualizations are rooted in a philosophical and cultural context and have significance beyond the data itself, so we should not only evaluate them in terms of functionality but should also consider the philosophical and aesthetic aspects of visualizations as cultural carriers.

    \section{Usage Scenarios}
\label{sec:usage-scenarios}

In this section, we discuss usage scenarios of \datasetName.
The usage scenarios are adapted from OldVisOnline~\cite{Zhang2024OldVisOnline} and contextualized with historical Chinese visualizations from \datasetName.
We also examine unique opportunities for culture-focused design.
Note that the scenarios below are not mutually exclusive.

\subsection{Searching Visualization}

\datasetName Gallery (\cref{fig:gallery}) allows efficiently searching historical Chinese visualizations.
Consider a scenario where a researcher is reading a paper discussing tables in \ChineseTerm{\Shiji}{史記}{Records of the Grand Historian} while the images provided in the original paper are not sufficient for understanding.
These visualizations cannot easily be queried on general-purpose search engines (e.g., Google Image Search). 
\revised{Thus, researchers have to find digital versions of these books and manually look through many pages to find the visualization they want, which is time-consuming and frustrating.}
With \datasetName Gallery, the researcher can directly search for these visualizations with a book name, such as \term{Shiji}. 
The researcher may also find different versions of a book.
In addition, similarity-based recommendations can help the researcher find more similar visualizations appearing in books other than the ones being queried to provide additional context.
Searching supported by \datasetName Gallery is also vital for other scenarios discussed below.

\subsection{Textual Criticism}

Textual criticism is a scholarly discipline that studies texts, primarily ancient documents, to reconstruct the original version.
\revised{As a dataset of historical visualizations, \datasetName may be used for textual criticism, as described by Zhang et al.~\cite{Zhang2024OldVisOnline}.}

\begin{figure}[!t]
    \centering
    \includegraphics[width=\linewidth]{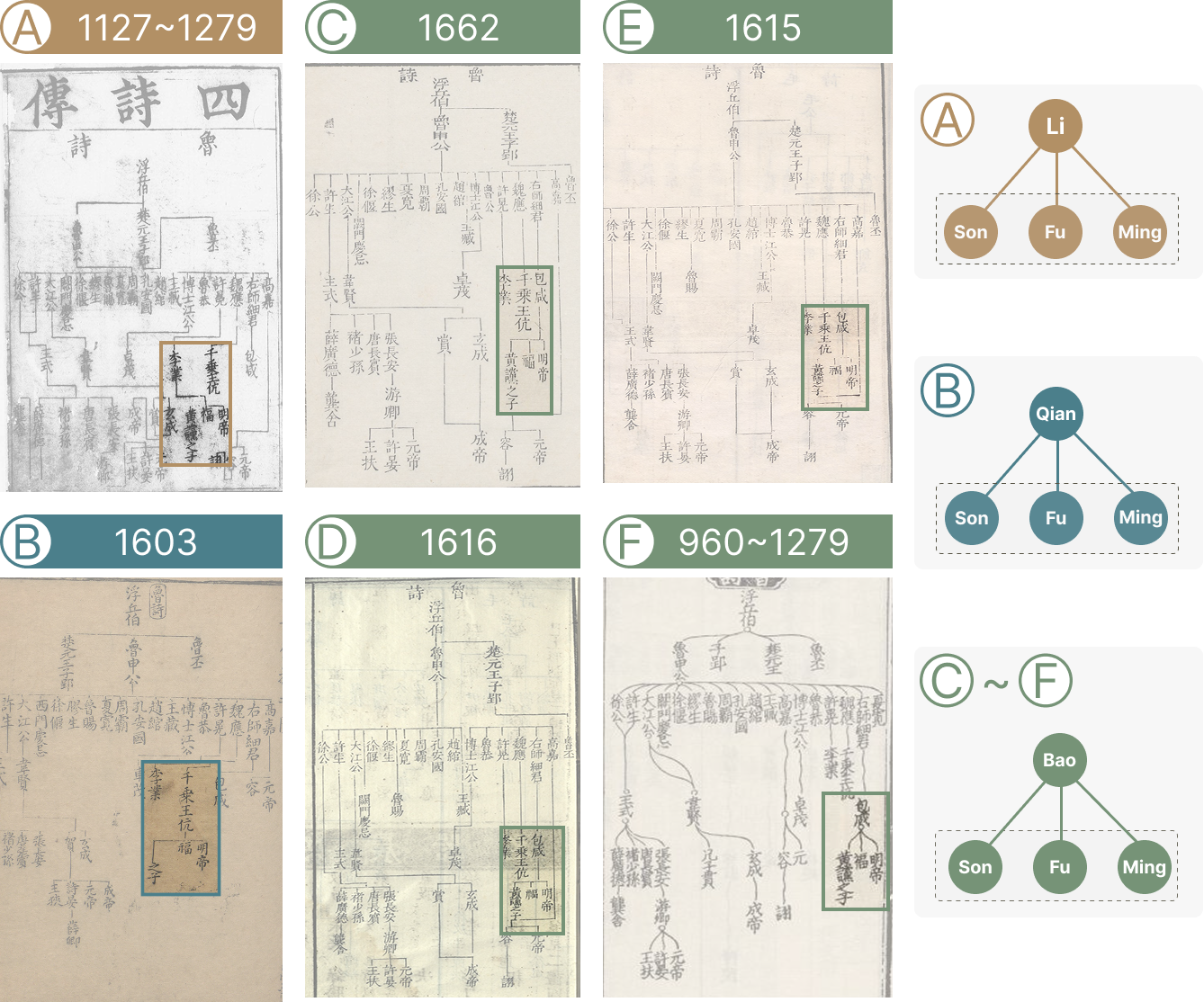}
    \caption{
        \revised{
        \textbf{Comparing editions of the \term{Lu} apprenticeship tree in \term{\SiShiChuanShouTu}:}
 Among the six editions, there are three variants of the apprenticeship relation of three apprentices: \term{Son}, \term{Fu}, and \term{Ming}.
        \href{https://commons.wikimedia.org/wiki/File:SBL003_毛詩擧要圖.pdf}{(A)},
        \href{https://tile.loc.gov/image-services/iiif/service:asian:lcnclscd:2014514341:2014514341v006:1E001-18b19a/full/pct:100/0/default.jpg}{(B)},
 and
        \href{https://iiif.lib.harvard.edu/manifests/view/drs:17209675$183i}{(C)}
        \href{https://digicoll.lib.berkeley.edu/record/73580?ln=en&v=pdf}{(D)}
        \href{https://www.digital.archives.go.jp/DAS/meta/listPhoto?LANG=default&BID=F1000000000000094161&ID=&NO=3&TYPE=JPEG&DL_TYPE=pdf}{(E)}
        \href{https://commons.wikimedia.org/wiki/File:ZHSY000019_監本纂圖重言重意互注點校毛詩二十卷圖譜一卷 (漢)毛萇 傳(漢)鄭玄 箋(唐)陸德明 釋文 宋刻本.pdf}{(F)}
 shows the master teaching the three apprentices as \term{Li}, \term{Qian}, and \term{Bao}, respectively.
 The published year (or estimated range) from the data sources is shown for each edition.
 }
 }
    \Description{
 Comparing six editions of the \term{Lu} apprenticeship diagram.
 }
    \label{fig:textual-criticism}
\end{figure}

Take \ChineseTerm{\SiShiChuanShouTu}{四詩傳授圖}{The Apprenticeships of the Four Schools of \ShiJing} as an example.
\revised{
It records the master-apprentice relations of four schools of \term{\ShiJing}.
Here, we focus on one of the schools: \term{Lu}.
\Cref{fig:textual-criticism} shows the \term{Lu} apprenticeship relations in six editions of \term{\SiShiChuanShouTu} stored in \datasetName.
By examining the structure of these tree visualizations, we observe three variants of a branch in the tree, which are highlighted in \cref{fig:textual-criticism}.
Specifically, the master who taught \term{Son}, \term{Fu}, and \term{Ming} was presented differently as \term{Li} (\cref{fig:textual-criticism}(A)), \term{Qian} (\cref{fig:textual-criticism}(B)), and \term{Bao} (\cref{fig:textual-criticism}(C - F)).
A transcription error may cause the differences, which is common during the compilation of ancient books.
Such cases are worth further investigation by historians.
}

\subsection{Investigating the History of Visualization}

\revised{
An understanding of the history of visualization is essential for its future development.
Due to the constraints on data retrieval and cultural barriers, as mentioned in \cref{sec:related-work}, the current study of visualization history is focused on the Eurocentric view.
Expanding our horizons to historical visualizations from different cultural contexts can help us understand visualization history more comprehensively and unbiasedly.
For example, when tracing the origin of tables, we observe Sima Qian's chronological tables in \term{\Shiji} created during the Western Han Dynasty, around 90 BCE (discussed in \cref{sec:taxonomy}).
Meanwhile, the origin of the formal chronology was previously attributed to Eusebius' \term{Chronicon}~\cite{Marchese2011Exploring}, which was created around 311 CE and similar to the vertical tables in \term{\Shiji}.
}

\subsection{Revitalizing Historical Graphic Designs}

\begin{figure}[!htb]
    \centering
    \includegraphics[width=\linewidth]{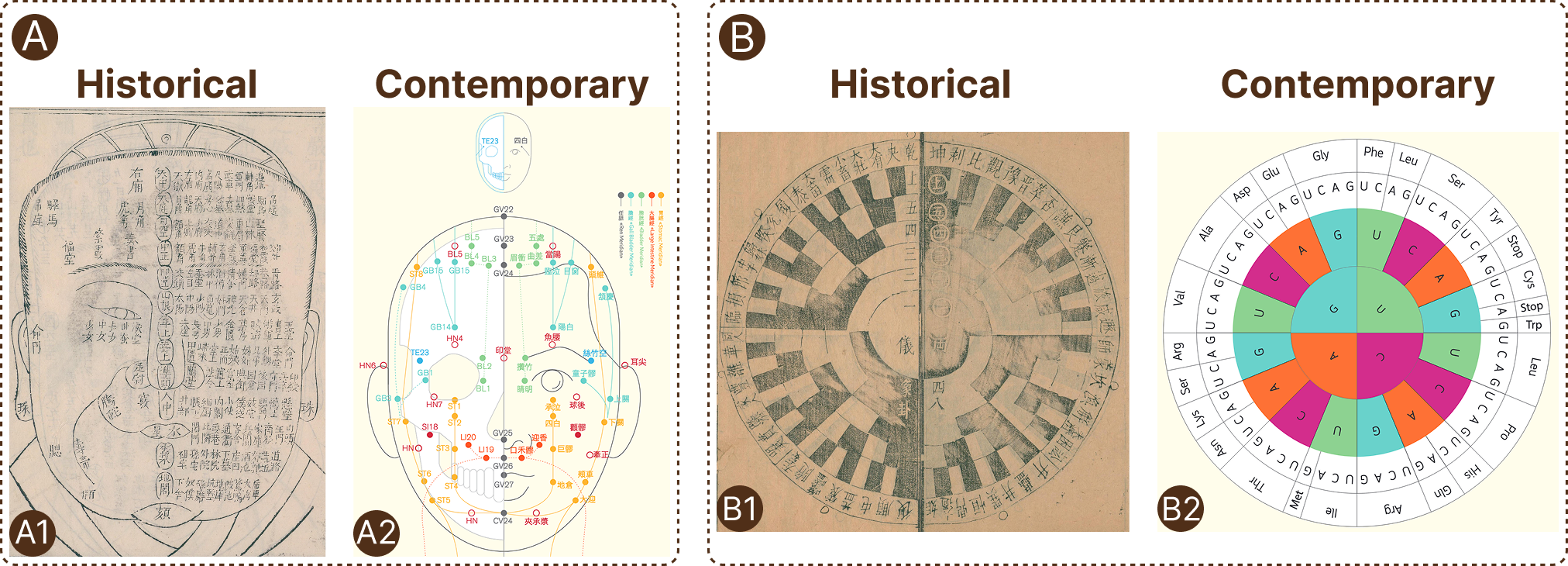}
    \caption{
 Two examples of contemporary designs inspired by historical designs created by Baur and Felsing~\cite{Baur2020Visual}:
 (A1) The \href{https://dl.ndl.go.jp/api/iiif/2574383/R0000010/full/full/0/default.jpg}{historical design} showing facial acupuncture points is from \term{San Cai Tu Hui}~\cite{Wang1609San}, 1609.
 (A2) The contemporary design incorporates smoother lines and combines visualization of acupuncture points with skeletal structures.
 (B1) A tree visualization of 64 hexagrams, stitched by the \href{https://dl.ndl.go.jp/api/iiif/2596442/R0000072/full/full/0/default.jpg}{left part}
 and the \href{https://dl.ndl.go.jp/api/iiif/2596442/R0000071/full/full/0/default.jpg}{right part}, from \term{Tu shu bian}~\cite{Huang1613Tu}, 1613.
 (B2) The contemporary design applies this visual encoding to map genetic codes to amino acids.
 }
    \Description{
 Two examples of contemporary designs inspired by historical designs created by Baur and Felsing.
 }
    \label{fig:redesign}
\end{figure}

Understanding and adapting visuals from history can facilitate design innovation and evoke audiences to rethink the role of contemporary designs.

\subsubsection{Inspire Design Innovation for Designers}
\label{sec:inspire-design-innovation}

By blending traditional visuals with modern design elements, designers may create unique expressions that respect cultural heritage while still appealing to contemporary tastes. 
Here, we introduce two interesting examples of such cross-cultural information designs created by Baur and Felsing~\cite{Baur2020Visual}.

\Cref{fig:redesign}(A) shows visualizations of facial acupuncture points.
\revised{In the historical design (\cref{fig:redesign}(A1)), two different styles of depiction are presented in the face: the left side of the face portrays pictorial facial features, while the right side is filled with extensive text annotations.}
In the contemporary design (\cref{fig:redesign}(A2)), the human face is also divided into two halves. 
The left side follows an anatomical style showing underlying skeletal structures, while the right side retains the historical design.
The contemporary design also uses color coding to distinguish different types of acupuncture points.

\Cref{fig:redesign}(B) demonstrates parallels between the structure of the 64 hexagrams derived from \term{Yin yang} and an amino acid codon table.
Both utilize a radial structure that expands outward from the center, with each layer progressively dividing until the outermost layer represents the hexagrams and the genetic codons, respectively~\cite{Baur2020Visual}.
Encoding codons based on the structure of the hexagram diagram facilitates the translation of a genetic code into an amino acid sequence.
\revised{The example indicates the potential of retargeting historical visualization designs to contemporary data.}

\subsubsection{Inspire Design Rethink for Audience}

Defamiliarization originates as a literary technique and serves as a method to question our habitual interpretations of everyday things~\cite{Bell2005Making}. 
In interface design, defamiliarization refers to altering familiar interfaces, interaction methods, or elements to encourage users to rethink and re-experience everyday interactions~\cite{Bell2005Making}.
\revised{Designers may leverage historical visual conventions to redesign contemporary visual elements and create a sense of visual defamiliarization for audiences}.
Visual defamiliarization, which involves presenting data through unfamiliar visual conventions, may encourage the audience to rethink the functional and cultural significance of contemporary design and help preserve historical visual conventions.
\revised{
Note that directly presenting unfamiliar visual conventions can lead to a cognitive burden on the audience, as discussed in~\cref{sec:discuss-convention}.
}

\subsection{Facilitating Culture-Focused Design}
\label{sec:culture-focused-design}

\revised{
Approaches to visual language can be placed on a continuum, according to Kostelnick~\cite{Kostelnick1995Cultural}, with the global approach at one end and the culture-focused approach at the other.
The global approach seeks to bridge the cultural gap by creating a rational, objective, and culturally neutral visual language.
The culture-focused approach emphasizes the strong correlation between visual communication effectiveness and cultural context, advocating for designs tailored to a specific cultural background.

While the global approach has become the guiding force in visual communication, culture-focused design deserves more attention.
Though they may be seen as niche or outdated, these designs emerging from specific historical and cultural contexts embody significant cultural value. 
We emphasize these design minorities to enrich the design diversity and promote design equity~\cite{Kostelnick2003Shaping}.
These designs can benefit certain situations, as users' cultural background can influence how they respond to information designs~\cite{Kostelnick1995Cultural}.

\datasetName can serve as a source of design materials to tailor designs for historical Chinese culture.}
The following lists two potential benefits of culture-focused designs suggested by Kostelnick and Hassett~\cite{Kostelnick2003Shaping}.

\begin{itemize}[leftmargin=3.5mm]
    \item \textbf{\revised{Streamline} comprehension:}
    \revised{Designs tailored for a specific culture can streamline the visual comprehension of those living in that culture~\cite{Kostelnick2003Shaping}.}
    Designers may leverage symbols and styles of historical Chinese graphics that encapsulate shared cultural knowledge and values to communicate more effectively with Chinese audiences.
    
    \item \textbf{Evoke emotional connotations:}
    \revised{Culture-focused designs can associate audiences with their cultural backgrounds to evoke emotional connotations.}
Using culturally specific visuals may lend credibility and authenticity to a design, positioning it within a particular tradition or discourse community, which can be beneficial on certain occasions~\cite{Kostelnick2003Shaping}.
    A commercial example discussed by~\cite{Kostelnick2003Shaping} is the cover design of ``The Old Farmer's Almanac'', the oldest continuously published periodical in North America. 
    Its cover has retained nearly two centuries of consistent visual features, including historical rural illustrations and vintage English typography~\cite{Kostelnick2003Shaping}. 
    These visual elements naturally connect readers to the country's profound natural and historical narrative, resonating deeply with its citizens.
\end{itemize}

\subsection{Games, Education, and Storytelling}

Historical Chinese \itemTypePlural and their redesigns can be combined with modern technologies to serve games, education, and storytelling.
Animation and interactivity can be applied to historical static visualizations to guide users' attention, enhancing their understanding of visualization encoding and historical contexts. 
These interactive ancient visualizations may even serve as serious games to support traditional cultural storytelling. 
For instance, the Xuanjitu project~\cite{RorySaur2022Xuanjitu} uses cyclic animations to display the positions of various poems within its palindrome structure, helping readers interpret historical literary works.
Moreover, historical Chinese visualizations contain rich visual expressions of cultural narratives, which may serve as resources for students to create their own historical data stories, thereby increasing classroom engagement and immersion~\cite{Lu2011ShadowStory}.
In recent years, more games such as Black Myth: Wukong~\cite{GameScience2024Black} have incorporated traditional cultural heritage into their scene and character designs. 
Graphics in \datasetName, such as for ancient clothing, architectural features, and perspective landscape paintings, may be used to train texture generation models for generating resources to be used in games.

    \section{Discussion}
\label{sec:discussion}

\revised{
In \cref{sec:lesson}, we reflect on our data curation process.
In \cref{sec:discuss-convention}, we reflect on the cross-cultural visualizations in terms of their formation, value, and guidelines for practical usage.
}

\subsection{\revised{Reflections on Data Curation}}
\label{sec:lesson}

We discuss some of our considerations and challenges in the data curation process.
We also discuss future work to extend \datasetName.

\textbf{Number of classes in detection:}
We set the object detection model to detect two classes, \taxon{visualization} and \taxon{illustration}, based on the following considerations.
We differentiate the two classes as our primary focus is on visualizations.
Meanwhile, the boundary between visualizations and illustrations is not always clear.
We include illustrations in the dataset so that the border can be easily revised during iterations, compared with not including illustrations at the time of detection.
Similarly, we categorize visualization into multiple subcategories after obtaining the visualization images.
While an alternative approach is to set the goal to directly detect images belonging to these subcategories, this alternative loses the flexibility of iteratively revising the boundary of visualizations.

\textbf{A label management strategy in classification:}
A strategy that may improve efficiency and quality of developing taxonomy labels in large scale data curation is to cluster the images that are extremely similar in visual appearance and constrain them to the same taxon.
For example, the images corresponding to different editions of a same visualization should be clustered.
Our consideration is that developing a taxonomy is an iterative process that requires frequent modifications.
By clustering similar images, we can ensure that the label modifications are applied to all images in the same cluster.
For large datasets, this approach may save time and reduce the risk of missing some images with labels to be edited when revising the taxonomy labels.
While we did not implemented this strategy in our own data curation process, looking back, we believe it could have been beneficial.

\textbf{Challenge in defining taxa boundaries for the taxonomy:}
In this work, we first grouped graphics into \taxon{visualization} and \taxon{illustration} and then further classified them based on their visual appearance.
As is the case for other taxonomy development processes~\cite{Chen2024Image}, clearly defining the boundaries of taxa is challenging.
A particular challenge for developing a taxonomy of historical Chinese visualizations is that due to textualism and pictorialism, the boundaries between visualization, text, and illustration can be blurred.
Additionally, the level of understanding of the historical artifacts may also influence that judgment.
To improve the current taxonomy call for expertise from not only visualization experts but also historians.

\textbf{Future work:}
Building a dataset of visualizations from diverse cultural frameworks is a gradual process. 
We believe that in the future, the dataset's content and label quality can be further improved in the following aspects.
First, classifying visualizations within a specific culture requires domain knowledge.
We may involve more domain experts to refine the taxonomy.
We may also investigate using vision-language model to assist in the classification of historical Chinese visualizations~\cite{Springstein2024Visual}.
So far, our data source is limited to three digital libraries.
The timeline in \cref{fig:gallery} shows that most of the books collected currently are from the Ming dynasty and later.
To expand the data coverage for earlier periods, we plan to utilize additional data sources, such as the National Library of China.
We plan to continue to expand the coverage of the \datasetName, and explore more applications to revive historical Chinese visualizations.

\subsection{Reflections on Cross-Cultural Visualization}
\label{sec:discuss-convention}

\begin{figure}[!htb]
    \centering
    \includegraphics[width=\linewidth]{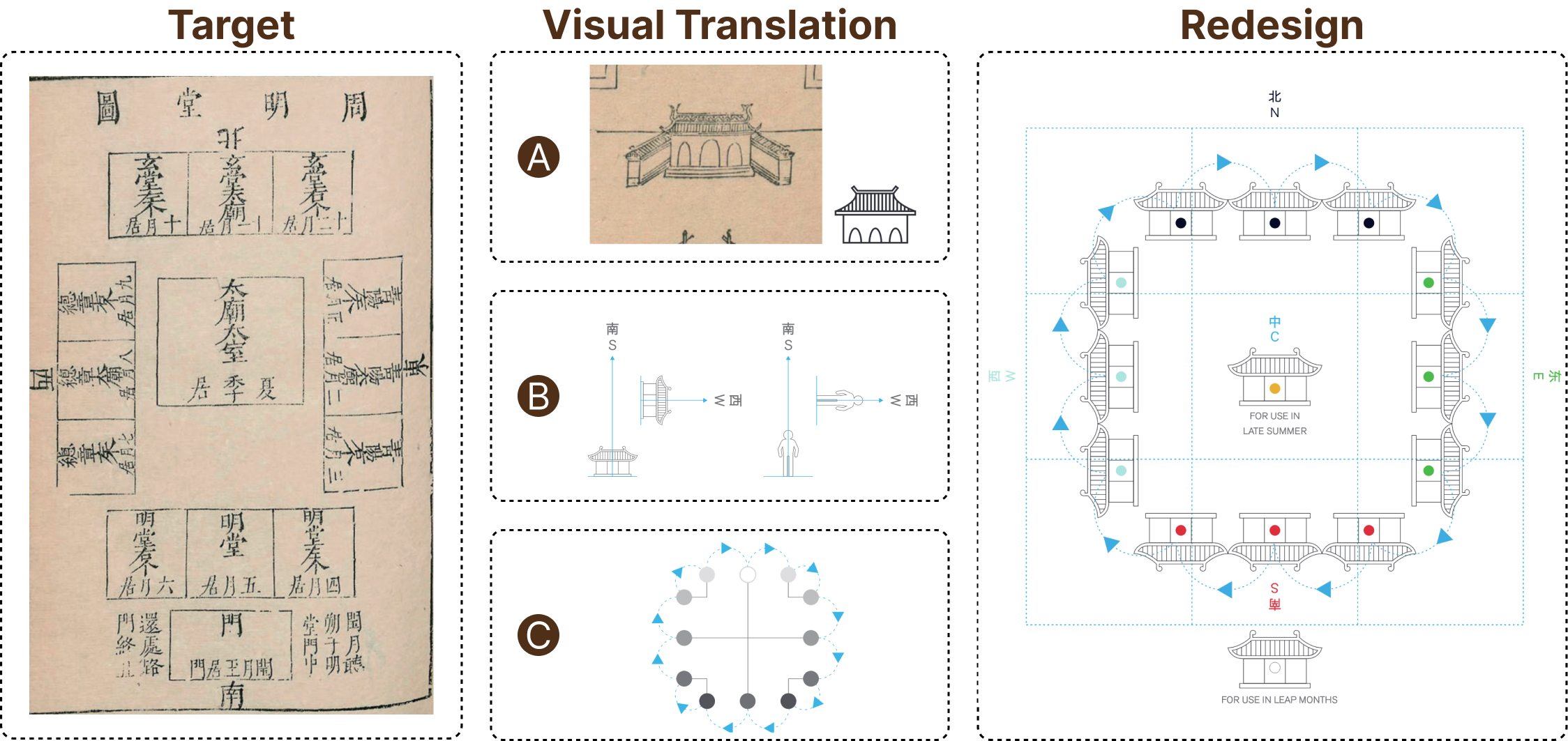}
    \caption{
        An example of visual translate \revised{created by Baur and Felsing}~\cite{Baur2020Visual}.
        The target visualization depicts different buildings occupied by ancient emperors during different time periods. 
        The redesign involves three visual translations.
        (A) translate rectangular building codes into ancient-style glyphs
        (B) a legend mapping building orientation to direction
        (C) arrows encode the emperor's movement trajectory over time
        The visualization is visually translated, making it more understandable to a wider audience.
    }
    \Description{
        An example of visual translation.
        The visualization represents ancient emperors' buildings and movements over time using glyphs for buildings, a legend for directions, arrows for movement, and colored circles for seasons, making it accessible to a broader audience.
    }
    \label{fig:translate}
\end{figure}

In this section, we use the concept of visual convention~\cite{Kostelnick2003Shaping} to reflect visualizations across different cultures from a higher-level perspective.
We discuss the formation and evolution of visual conventions across communities and examine practices for promoting them in contemporary applications.

\textbf{Definition of visual conventions:}
A discourse community is a group of people with shared goals, practices, and knowledge, especially in communicating.
These groups establish and use shared visual conventions. 
A visual convention is a system of shared symbols or practices that people recognize over time, enabling clear and efficient visual communication~\cite{Kostelnick2003Shaping}.
For example, musicians form a discourse community, and musical notation is one of their visual conventions~\cite{Kostelnick2003Shaping}.

\textbf{Impact factors for visual conventions:}
Visual conventions change due to technology advancements and shifts in sociocultural factors~\cite{Kostelnick2003Shaping}.
\revised{
As new communities form and old ones disappear, some conventions are forgotten over time, while others last and grow in influence.
On one hand, advancements in modern technology, such as web-based visualization, have shaped new contemporary visual conventions.
For example, in an interactive visualization, when a user hovers over an element, the de-highlighted elements are considered irrelevant to the currently hovered element. In contrast, the highlighted elements are considered relevant.
On the other hand, it also hindered the continuity of ancient conventions. 
Visual elements rendered in HTML are mostly restricted to fluent modernist lines, losing the historical hand-drawn style.
}
Cultural and organizational contexts also shape distinct visual conventions across nations and institutions~\cite{Kostelnick2003Shaping}. 
One potential explanation arises from Hall's theory about high-context and low-context culture~\cite{Hall1977Culture}. 
In high-context cultures (e.g., China, Korea, Japan~\cite{Kim1998High}), communication tends to be implicit, relying heavily on background context, social relationships, and environmental cues. 
This emphasis on context is reflected in design, where metaphor and symbolism are often employed to convey deeper meanings. 
In contrast, low-context cultures (e.g., America~\cite{Kim1998High}) rely on explicit information encoding with minimal dependence on context. 
This is reflected in designs that prioritize clarity and directness.

\textbf{Practice for utilizing visual conventions:}
Designers need to ensure that the audience can understand visual conventions from different cultures or histories when using them in design.
On one hand, as discussed in~\cref{sec:culture-focused-design}, it can streamline comprehension for those who live under that cultural framework and evoke a rethink of contemporary designs.
On the other hand, if designs are presented directly with unfamiliar visual conventions without any explanation, the audience may not understand the designs and become confused and frustrated.
Therefore, information designers should provide a ``visual translation''~\cite{Baur2019Cultural} to bridge that gap between different visual conventions, as well as maintain the characteristics for them.
Visual translation involves reinterpreting culturally specific visual designs and conventions using modern, universally recognized visual norms~\cite{Baur2019Cultural}. 
Critical cultural information embedded in the original design may not be visually apparent or understood by outsiders but can be conveyed through shared visual norms and appended to the original design.
\Cref{fig:translate} shows an example designed by Baur and Felsing~\cite{Baur2020Visual}.
It employs three visual translators (\cref{fig:translate}(A), \cref{fig:translate}(B), and \cref{fig:translate}(C)) to reveal the implicit knowledge embedded in historical Chinese layout plan visualizations.
\Cref{fig:translate} depicts the \term{Mingtang}, an ancient ceremonial building symbolizing cosmic order through its architecture.
Its layout guided the emperor's ceremonial activities, and each room corresponds to a specific month, determining where the emperor should be during different months of the year.
Three visual translations are employed to convey this information.
\Cref{fig:translate}(A) uses a glyph extracted from traditional Chinese architecture to replace the original rectangular encoding, which conveys ancient aesthetics and reduces cognitive load. 
\Cref{fig:translate}(B) acts as a legend indicating the relationship between the direction and buildings' orientation. 
\Cref{fig:translate}(C) uses a circular arrangement with sequential arrows to imply the function of guiding emperor activities.
Finally, these visual translations are integrated with the original visualizations, incorporating color-coded circles to symbolize the seasons, forming the final translated visualization. 
The new visualization preserves the historical Chinese visual conventions while aligning them with the visual conventions in modern society to improve understanding across different cultures.

    \section{Conclusion}

We introduced \datasetName, a dataset of historical Chinese visualizations and illustrations.
To collect these graphics, we implemented a semi-automatic pipeline by which we extracted \numVisShort visualizations and \numIllusShort illustrations from historical Chinese books.
Based on \datasetName, we combine historical factors to analyze and explain design patterns in historical Chinese visualizations.
Our analysis highlighted the distinct features of historical Chinese visualizations, such as textualism and pictorialism.
We also envision the potential usage scenarios of \datasetName, including supporting textual criticism, and \revised{inspiring culture-focused designs.}
\revised{
Through our effort in curating the dataset, we aim to draw public attention to facilitate a basic understanding of historical Chinese visualizations.
Ancient Chinese culture is, however, only one of the underrepresented cultural communities in the common narrative of visualization history.
We call for more investigations into historical visualizations under other underrepresented cultures, which are critical for a comprehensive and unbiased understanding of the history of visualization.
}

\begin{acks}
This work was supported by NSFC No. 62272012. It is also partially supported by Wuhan East Lake High-Tech Development Zone National Comprehensive Experimental Base for Governance of Intelligent Society.
\end{acks}

\bibliographystyle{ACM-Reference-Format}
\bibliography{assets/bibs/papers,assets/bibs/historical-visualizations,assets/bibs/custom}

\end{document}